\documentclass[%
 reprint,
preprintnumbers,
nofootinbib,
 amsmath,amssymb,
 aps,
 prd,
]{revtex4-2}

\usepackage{times} %
\clearpage{}%

\newcommand*{\LIGODCCNumber}{P1900058}
\newcommand\this{paper}
\newcommand\lensedhyp{\mathcal{H}_{\rm L}}
\newcommand\notlensedhyp{\mathcal{H}_{\rm NL}}
\newcommand{\boldvec}[1]{\bm{#1}}
\newcommand\diff{\mathrm{d}}
\newcommand\ppop{p_{\rm pop}}
\newcommand\srcpop{p_{\rm src}}
\newcommand\lenspop{p_{\rm lens}}

\newcommand{\comparam}[1]{\bm{\theta}^{#1}_{\rm{com}}}
\newcommand{\idparam}[1]{\bm{\theta}^{#1}_{\rm{ind}}}
\newcommand\pcom{p_{\rm pop, com}}
\newcommand\pid{p_{\rm pop, ind}}
\newcommand{\bayesfactor}[2]{\mathcal{B}^{#1}_{#2}}
\newcommand{\posteriorodds}[2]{\mathcal{O}^{#1}_{#2}}
\newcommand{\priorodds}[2]{\mathcal{P}^{#1}_{#2}}
\newcommand\cohratio{\mathcal{C}}
\newcommand\Ntrig{N}
\newcommand\Nsrc{N_{\rm src}}
\newcommand\data{D}
\newcommand\popparam{\bm{\lambda}}
\newcommand\lensparam{\bm{\gamma}}
\newcommand\eventparam{\theta}
\newcommand\pext{p_{\rm ext}}\clearpage{}%

\usepackage{glossaries}
\clearpage{}%
\newacronym{GW}{GW}{gravitational-wave}
\newacronym{EM}{EM}{electromagnetic-wave}
\newacronym{LIGO}{LIGO}{Laser Interferometer Gravitational-wave Observatory}
\newacronym{CBC}{CBC}{compact binary coalescence}
\newacronym{BBH}{BBH}{binary black hole}
\newacronym{PE}{PE}{parameter estimation}
\newacronym{IMR}{IMR}{inspiral-merger-ringdown}
\newacronym{MLE}{MLE}{maximum likelihood estimator}
\newacronym{MAP}{MAP}{maximum a posteriori estimator}
\newacronym{O1}{O1}{the first observing run}
\newacronym{O2}{O2}{the second observing run}
\newacronym{aLIGO}{aLIGO}{Advanced LIGO}\clearpage{}%

\usepackage{graphicx}%
\graphicspath{{./}{figures/}}
\usepackage[caption=false]{subfig}
\usepackage{ifthen}
\usepackage{tikz}
\usetikzlibrary{bayesnet}
\usetikzlibrary{matrix,calc}
\usepackage{xcolor}

\usepackage{algorithm}
\usepackage{algpseudocode}

\usepackage{dcolumn}%
\usepackage{multirow}
\usepackage{amsmath,amsfonts,amssymb,bm}
\usepackage{bm}%
\usepackage{hyperref}%

\begin{document}

\title{Bayesian statistical framework for identifying \\strongly lensed gravitational-wave signals}

\author{Rico K.~L.~Lo}
\email{kllo@caltech.edu}
\affiliation{LIGO, California Institute of Technology, Pasadena, California 91125, USA}

\author{Ignacio Maga\~na Hernandez}
\affiliation{University of Wisconsin-Milwaukee, Milwaukee, Wisconsin 53201, USA}

\date{\today}%

\begin{abstract}
It is expected that gravitational waves, similar to electromagnetic waves, can be gravitationally lensed by intervening matters, producing multiple instances of the same signal arriving at different times from different apparent luminosity distances with different phase shifts compared to the unlensed signal due to lensing. If unaccounted for, these lensed signals will masquerade as separate systems with higher mass and lower redshift. 
Here we present a Bayesian statistical framework for identifying strongly lensed gravitational-wave signals that incorporates astrophysical information and accounts for selection effects. We also propose a two-step hierarchical analysis for more efficient computations of the probabilities and inferences of source parameters free from shifts introduced by lensing.
We show with examples on how changing the astrophysical models could shift one's interpretation on the origin of the observed gravitational waves, and possibly lead to indisputable evidence of strong lensing of the observed waves. In addition, we demonstrate the improvement in the sky localization of the source of the lensed signals, and in some cases the identification of the Morse indices of the lensed signals.
If confirmed, lensed gravitational waves will allow us to probe the Universe at higher redshift, and to constrain the polarization contents of the waves with fewer detectors. \end{abstract}

\maketitle

\section{Introduction}
\label{sec:introduction}

As gravitational waves propagate through the Universe to the Earth, they can be deflected, or lensed, by intervening matters such as galaxies or galaxy clusters acting as gravitational lenses, just like electromagnetic waves. For visible light coming from a distant background source, in the case of strong lensing where the deflection is sufficiently large, multiple images of the source will be formed that are close to each other, typically separated by only several arcseconds and distorted compared to the unlensed image \cite{Schneider}. For transients, the time variation of the images are correlated and delayed, where the time delays range from days to months \cite{Schneider}. For transient \gls{GW} signals, such as those emitted from coalescences of compact binary systems, multiple images refer to multiple \gls{GW} triggers registered at different times by \gls{GW} detectors.\footnote{We use the term image and signal interchangeably.}
For an in-depth review on strong lensing of explosive transients across the \gls{EM} and the \gls{GW} spectra, see for example Ref.~\cite{Oguri:2019fix}.

To investigate the effects of strong lensing on a \gls{GW} signal, we first briefly describe how the observed GW strain $h(t)$ depends on some parameters
$\bm{\eventparam}$ when the signal is not affected by strong lensing.\footnote{Here we use $\bm{\eventparam}$ as a placeholder to denote any generic set of parameters, which can consist of different parameters in different contexts.} The \gls{GW} strain observed by a laser-interferometric GW detector is given by
\begin{equation}
\label{eq:observed_strain}
h(t - t_{\rm c}; \bm{\eventparam}) = \frac{1}{d_{\rm L}(z)} \sum_{\mathrm{pol} = +,\times}F_{\mathrm{pol}}(\alpha, \delta, \psi; t_{\rm c}) h_{\mathrm{pol}}(t - t_{\rm c}; \bm{\eventparam}),
\end{equation}
where the detected GW strain is a projection along the arms of the detector where the response of the detector to the two polarization states of the GW is defined by the detector's beam pattern functions $F_{+, \times}(\alpha, \delta, \psi)$, where $\alpha$ is the right ascension, $\delta$ is the declination, and $\psi$ is the polarization angle of the source respectively.\footnote{Note that the beam pattern functions $F_{+, \times}$ are usually calculated in a frame where the detector is situated at the origin, where the angles $\alpha$ and $\delta$ are celestial coordinates. Therefore, the beam pattern functions depend implicitly on the time of the event as well.} The luminosity distance to the source $d_{\rm L} (z)$ is a function of the redshift $z$ and depends explicitly on the cosmology. The time and phase at coalescence of the signal are denoted by $t_{\rm c}$ and $\phi_{\rm c}$ respectively. The waveform of the two polarization states $h_{+}$ and $h_{\times}$ can be compactly written as
\begin{equation}\label{eq:GW_waveform}
\begin{aligned}
	h_{+}(t; \bm{\eventparam}) - ih_{\times}(t; \bm{\eventparam}) = \sum_{\ell=2}^{\infty} \sum_{m=-\ell}^{\ell} h_{\ell m}\left(t ; \bm{\eventparam} \right) {}_{-2}Y_{\ell m} \left( \iota, \phi_{o} \right),
\end{aligned}
\end{equation}
where $h_{\ell m}$ depends on some intrinsic parameters $\vartheta$ and the source redshift $z$ only, and ${}_{-2}Y_{\ell m} \left( \iota, \phi_{o} \right)$ is the spin-weighted spherical harmonic that depends on the polar angle $\iota$ and the azimuthal angle $\phi_{o}$ in the source frame.\footnote{For nonprecessing binary systems, the angle $\iota$ is also known as the inclination angle, the angle between the line of sight and the orbital angular momentum vector, which is by convention chosen to be along the $z$-axis of the source frame. For generic precessing binary systems, the inclination angle changes over time as the orbital angular momentum vector precesses around the total angular momentum vector.}\footnote{Note that the phase at coalescence $\phi_{\rm c}$ does not enter Eq.~\eqref{eq:GW_waveform} explicitly but through the time-varying azimuthal angle $\phi_{o}$, with $\phi_{\rm c}$ as the reference phase.}

For the case of \glspl{GW} from quasicircular \gls{BBH} mergers, the set of intrinsic parameters $\vartheta =  \{ \mathcal{M}_{\rm c}, q, \boldvec{\chi}_1, \boldvec{\chi}_2 \}$ where $\mathcal{M}_{\rm c} \equiv (m_1 m_2)^{3/5}/(m_1 + m_2)^{1/5}$ is the chirp mass of the binary, $q \equiv m_2/m_1 \leq 1$ is its mass ratio, both in terms of the binary component masses $m_1 \geq m_2$. The vectors $\boldvec{\chi}_1$, $\boldvec{\chi}_2$ are the dimensionless spin vectors for the binary components. Note that this is just one particular parametrization, and other ways of specifying the intrinsic parameters are also possible; for example, using the total mass of the binary $M_{\rm tot} \equiv m_1 + m_2$ instead of the chirp mass $\mathcal{M}_{\rm c}$. To account for the effect of an expanding universe, one can do so by simply replacing the (source-frame) masses $m_{i}^{\rm src}$ with the (detector-frame) redshifted masses $m_{i}^{\rm {det}} = (1 + z)m_i^{\rm {src}}$. Similarly, we define the redshifted chirp mass as $\mathcal{M}_{\rm c}^{\rm {det}} = (1 + z)\mathcal{M}_{\rm c}^{\rm {src}}$, while the mass ratio remains unchanged. It should be noted that Eqs. \eqref{eq:observed_strain} and \eqref{eq:GW_waveform} hold true regardless of the type of the binary system under consideration, and in general the set of intrinsic parameters $\vartheta$ will differ for each type of binary systems.

Working in the geometric optics limit where the wavelength is much shorter than the lens length scale, for a majority of the time, strongly lensed GW signals from a binary system will have the same morphology with different amplitudes (corresponding to different magnifications) and arrive at different times. A given image has an absolute magnification $\mu$ which can be defined in terms of the true luminosity distance to the source $d_{\rm L}^{\rm {src}}$ and the apparent luminosity distance $d_{\rm L}$ as
\begin{equation}
\label{eq:Apparent_luminosity_distance}
    d_{\rm L}^{(i)} = \frac{d_{\rm L}^{\rm {src}}}{\sqrt{\mu^{(i)}}},
\end{equation}
where the bracketed superscript indexes the images.
The lensed images arrive at the (center of the) Earth at different times because of the geometrical time delay, as they follow different null trajectories, and the time delay due the gravitational potential.
We define the relative time delay $\Delta t$ between two images as $\Delta t \equiv t_{\rm c}^{(2)} - t_{\rm c}^{(1)}$, where $t_{\rm c}$ is the GW arrival (trigger) time for each image. Here we assume that the time delay $\Delta t$ is large enough so that the lensed images will not overlap with each other and that we can identify them as separate triggers. We refer to the image that arrives first as the first image, and vice versa, such that $\Delta t > 0$. We can also define the relative magnification $\mu_{\rm rel}$, which is simply the ratio of the two absolute magnifications as
\begin{equation}
	\mu_{\rm rel} \equiv \frac{\mu^{(2)}}{\mu^{(1)}}.	
\end{equation}
The strong lensing of GW can also induce nontrivial effects on the GW waveform other than a change in the amplitude of the signal and a simple shift in time. In general, the lensed waveform $\tilde{h}_{\rm pol}^{\rm lensed}(f)$ is related to the unlensed waveform $\tilde{h}_{\rm pol}^{\rm unlensed}(f)$ by a frequency-dependent amplification factor $F(f)$ as \cite{Schneider, Takahashi:2003ix}
\begin{equation}
	\tilde{h}_{\rm pol}^{\rm lensed}(f) = F(f)\;\tilde{h}_{\rm pol}^{\rm unlensed}(f),
\end{equation}
for $f > 0$. The negative frequency components can be obtained via the reality condition $\tilde{h}_{\rm pol}(-f) = \tilde{h}_{\rm pol}^{*}(f)$. In the geometric optics limit, the amplification factor is given by \cite{Schneider,Takahashi:2003ix}
\begin{equation}
\label{eq:Phasing_effect}
	F(f) = \sum_{j} \sqrt{\mu^{(j)}} \exp(2\pi if\Delta t^{(j)} - in^{(j)} \pi/2),
\end{equation}
for $f>0$ and $n^{(j)} \in \left\{ 0, 1, 2\right\}$ is known as the Morse index of the $j$th image \cite{Schneider}.\footnote{Unlike Ref.~\cite{Dai:2017huk}, we did not include the $\mathrm{sgn}(f)$ factor explicitly in Eq.~\eqref{eq:Phasing_effect}. This is because by imposing the reality condition on $\tilde{h}_{\rm pol}(f)$ to obtain the negative frequency components, adding an additional $\mathrm{sgn}(f)$ will have no effect to the final expression for the waveform.} The factor $\sqrt{\mu}$ causes the apparent luminosity distance to differ from the luminosity distance of the source, and the phase factor $\exp(2\pi if\Delta t)$ causes the aforementioned time delay. The frequency-independent phase shift from $\exp(-in \pi/2)$ is degenerate with a shift in the phase at coalescence when we consider GW signals from nonprecessing binaries with contributions only from the quadrupole $\ell = |m| = 2$ modes \cite{Dai:2017huk, Ezquiaga:2020gdt}. When the geometric optics approximation breaks down and the full wave optics treatment is needed, the expression for the amplification factor can be much more complicated than Eq.~\eqref{eq:Phasing_effect}, and encodes more information about the gravitational lens \cite{Schneider, Takahashi:2003ix}.

Previous works have shown that the detection rate for lensed GWs could be, optimistically, $5^{+5}_{-3}\,\textrm{yr}^{-1}$ \cite{Ng:2017yiu}
for Advanced LIGO \cite{TheLIGOScientific:2014jea} and Advanced Virgo \cite{TheVirgo:2014hva} operating at their design sensitivities, while others predicted more pessimistic rates, ranging from $0.58 \,\textrm{yr}^{-1}$ \cite{Oguri2018} to $1.20 \,\textrm{yr}^{-1}$ \cite{Li:2018prc} depending on the source population model assumed. The detection rate for lensed GWs can also be constrained from the detection, and more surprisingly the nondetection, of stochastic gravitational-wave background from individually unresolvable binaries \cite{Buscicchio:2020cij, Mukherjee:2020tvr}. Searches on \gls{O1} and \gls{O2} data for strongly lensed GW signals were performed \cite{Hannuksela:2019kle,Dai:2020tpj,Liu:2020par}, and it was concluded that there is no significant evidence that any of the eleven detected GW events during O1 and O2 are lensed, while Ref.~\cite{2019arXiv190103190B} suggests that GW170809 and GW170814 could be lensed images of each other due to the similarity of the waveforms for these two events. 

It is also possible that strongly lensed GW signals from distant sources are absolutely de-magnified ($\mu < 1$), or relatively demagnified ($\mu_{\rm rel} < 1$)  such that when found by matched-filtering search pipelines (such as \texttt{GstLAL} \cite{Messick:2016aqy, Sachdev:2019vvd} and \texttt{PyCBC} \cite{Usman:2015kfa}) they appeared to be subthreshold triggers, not statistically significant enough to claim detections. If a lensed image is loud enough to be detected individually, then targeted matched-filtering based searches can be performed for each detected \gls{GW} event and search deeper for its potentially subthreshold lensed image counterparts \cite{Li:2019osa, McIsaac:2019use, Dai:2020tpj}.

An overarching approach in searches for lensed GW signals is the use of Bayesian statistics \cite{Haris:2018vmn, Hannuksela:2019kle, McIsaac:2019use, Liu:2020par, Dai:2020tpj}, where a statistic, either called a ``Bayes factor'' in the usual context of Bayesian hypothesis testing or a ranking score, is calculated.
In this \this, we present a Bayesian statistical framework for identifying strongly lensed \gls{GW} signals that utilizes hierarchical Bayesian modeling. By modeling the data generation processes when the observed GW signals are lensed and not lensed respectively, we develop a framework that allows us to compute a Bayes factor, and hence a posterior odds, that incorporates astrophysical information directly and accounts for selection effects.
We argue that in order to interpret the Bayes factor properly as a ratio of normalized probability densities of the observed data, selection effects cannot be ignored and must be accounted for in order to normalize the probability densities. The ability to directly incorporate astrophysical information, both on the GW sources as well as the gravitational lenses, serves to better bridge the astrophysical modeling community and the GW data analysis community. In addition, we argue that whether a GW signal is interpreted as lensed or not depends also on the astrophysical models assumed, making the prior astrophysical information an indispensable ingredient of the analysis.

The \this{} is structured as follows: Section~\ref{sec:statistical_framework} presents the hierarchical Bayesian framework for identifying strongly lensed \gls{GW} signals in a general setting, and the technique to marginalize over the source redshift separately and infer the true source parameters. In Sec.~\ref{sec:BBH_pair_lensing} and Sec.~\ref{sec:BBH_single_lensing} we apply and showcase the statistical framework to analyze strongly lensed \gls{GW} signals from \gls{BBH} mergers when we analyze two GW signals jointly and analyze one signal at a time respectively. Throughout the \this{}, we assume a flat $\Lambda$CDM cosmology with $H_{0} = 67.7 \; \rm{km}\,\rm{s}^{-1}\,\rm{Mpc}^{-1}$ and $\Omega_{\rm m} = 0.307$ from the Planck 2015 results \cite{Ade:2015xua}. 
\section{Statistical framework}
\label{sec:statistical_framework}
In order to differentiate strongly lensed \gls{GW} signals from \gls{GW} signals that are not lensed, we adopt a Bayesian statistical framework where we introduce two models/hypotheses\footnote{We will use the word model and hypothesis interchangeably throughout the \this{}.} that we want to compare; namely, the lensed hypothesis $\lensedhyp$ and the not-lensed hypothesis $\notlensedhyp$.\footnote{The not-lensed hypothesis $\notlensedhyp$ is often referred as the unlensed hypothesis, denoted by $\mathcal{H}_{\rm U}$, in literature. Here we reserve the meaning of unlensed for effects due to lensing being reverted.}
The framework applies for $\Ntrig \geq 1$, where $\Ntrig$ is the number of \gls{GW} events under consideration, unlike much of previous work \cite{Haris:2018vmn, Hannuksela:2019kle, McIsaac:2019use, Liu:2020par, Dai:2020tpj}.
In the following we will elaborate on what we mean by the two hypotheses, and model the data generation processes under the two different hypotheses using hierarchical Bayesian modeling.

\subsection{The not-lensed hypothesis $\notlensedhyp$}
\begin{figure}[h]
\begin{center}
\includegraphics[width=\columnwidth]{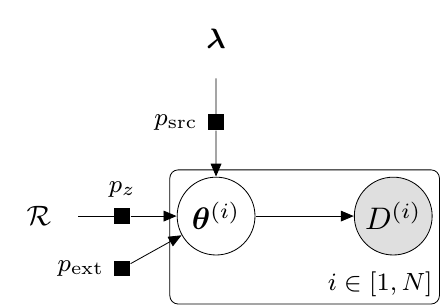}
\end{center}	
\caption{\label{fig:not_lensed_hyp_data_gen}Data generation process for the $\Ntrig$ observed data under the not-lensed hypothesis $\notlensedhyp$. Each data $\data^{(i)}$ can be described by the event-level parameters $\bm{\eventparam}^{(i)}$ which were drawn from the population-informed prior distribution $\ppop = \srcpop(\bm{\eventparam}^{(i)}|\bm{\popparam})p_{z}(z^{(i)}|\mathcal{R}, \notlensedhyp) \pext$ with $\popparam$ controlling the shape of the source population distribution, $\mathcal{R}$ being the merger rate density, and $\pext$ describing the distribution of the extrinsic parameters except for the redshift.}	
\end{figure}

Suppose we have $\Ntrig$ \gls{GW} events under consideration, i.e., we have a set of $\Ntrig$ time series data $\bm{\data} = \{ \data^{(i)}\}_{i=1}^{i=\Ntrig}$, where the bracketed superscript indexes the events. The not-lensed hypothesis means that the observed $\Ntrig$ events are $\Ntrig$ independent realizations of a population distribution of \gls{GW} source $\srcpop$, parametrized by some parameters $\boldvec{\bm{\popparam}}$ that control only the shape of the distribution and the total number of sources $\Nsrc$ in that population.\footnote{Note that the source population distribution $\srcpop(\bm{\eventparam}|\bm{\popparam})$ is normalized such that $\diff \Nsrc/\diff \bm{\eventparam} = 1/\Nsrc \; \srcpop(\bm{\eventparam}|\bm{\popparam})$.}
Note that we have assumed all $\Ntrig$ of them are of astrophysical origins.
Simply put, the $\Ntrig$ events are just $\Ntrig$ different systems, with the event-level parameters $\bm{\bm{\eventparam}}^{(i)}$ (such as component masses and spins) describing the $i$th event being randomly drawn from a source population distribution $\srcpop(\bm{\bm{\eventparam}}|\bm{\bm{\popparam}})$, where $\bm{\bm{\popparam}}$ might be for example the maximum mass of a black hole in that population. These $\Ntrig$ signals will have different source redshifts $z$ drawn from the distribution $p_{z}(z^{(i)}|\mathcal{R})$, where $\mathcal{R} = \mathcal{R}(z)$ is the merger rate density that can be a function of the source redshift $z$, and with different extrinsic parameters such as the sky location drawn from the distribution $\pext$. A concise way of expressing this is that $\bm{\eventparam}^{(i)} \sim \ppop(\bm{\eventparam}^{(i)})$ where $\ppop = \srcpop(\bm{\eventparam}^{(i)}|\bm{\popparam})p_{z}(z^{(i)}|\mathcal{R}, \notlensedhyp) \pext$ is the population-informed prior distribution under the not-lensed hypothesis. The event-level parameters $\bm{\eventparam}^{(i)}$ then in turn ``generate'' the data $\data^{(i)}$ that we observed for the $i$th event. Figure \ref{fig:not_lensed_hyp_data_gen} shows a graphical representation of this data generation process.
Although we are not making any inference on the population-level parameters of GW sources and instead we fix them in our analysis (i.e. choosing $\popparam$ and $\mathcal{R}$ a priori), we see that the problem of identifying strongly lensed signals can be naturally framed as a population analysis. Moreover, we can reuse many of the results from rates and population analyses (for example see Refs. \cite{LIGOScientific:2018jsj, Abbott:2020gyp}).

\subsection{The lensed hypothesis $\lensedhyp$}
For the lensed hypothesis, suppose we also have the same $\Ntrig$ events under consideration. However, the lensed hypothesis means that these $\Ntrig$ events are actually $\Ntrig$ strongly-lensed images of the same source. Instead of drawing $\Ntrig$ independent realizations from the population distribution $\srcpop$, now we only have one realization of this source population distribution as the images correspond to the same GW source. In addition to the source population distribution, we will need to introduce the lens population distribution $\lenspop$, parametrized by some parameters $\bm{\lensparam}$, that describes for example the joint probability distribution of the absolute magnification of lensed images. Furthermore, we partition the event-level parameters $\bm{\eventparam}^{(i)}$ into two disjoint sets: common parameters $\comparam{(i)}$ and independent parameters $\idparam{(i)}$. For the common parameters $\comparam{(i)}$ we expect them to be the same across the $\Ntrig$ signals, for example the masses and spins of the source binary system, as the $\Ntrig$ events correspond to the same source. In addition to the source parameters, we also expect the redshift $z^{(i)}$ of each image to be the same as the source redshift as strong lensing is achromatic, leaving the redshift unchanged. For extrinsic parameters, we can also assume them to be the same except for the (apparent) luminosity distance and the time of arrival. While it is true that strong lensing will deflect a \gls{GW} signal from its original null trajectory, the typical deflection angle for gravitational lensing due to a galaxy or a galaxy cluster is only of the order of arcseconds and arcminutes respectively \cite{Schneider, Singer:2019vjs}, which is much smaller than the typical uncertainty in the source localization of a GW signal. Therefore, it is valid to assume that the $\Ntrig$ images share the same sky location. We also expect the difference in the polarization angle $\psi$ to be negligible \cite{Hou:2019wdg}. In summary, the common parameters $\comparam{(i)}$ are one random draw of the distribution $\pcom = \srcpop(\comparam{(i)}|\bm{\popparam})p_{z}(z^{(i)}|\mathcal{R}, \lensedhyp) \pext$, where $\pcom$ is the population-informed prior for the common parameters $\comparam{}$ under the lensed hypothesis.

As for the independent parameters $\idparam{(i)}$, we expect them to be different for each event. For example, the absolute magnification $\mu$ and the arrival time $t_{\rm c}$ of each image would be different. Note that the dimension of the event-level parameters $\bm{\eventparam}$ under the lensed hypothesis can be different than that under the not-lensed hypothesis. For example, different lensed images can be classified into three types where each type of an image will have a different phasing effect to the lensed waveform [for example see Eq.~\eqref{eq:Phasing_effect}]. The number of lensed images produced by a gravitational lens can also inform us on the type of lens that produces the images. Here we do not use this information since it is possible (and often the case) that we are only analyzing a subset of lensed images coming from a particular source and lens, either deliberately or simply because we did not observe all of the lensed images. In short, each image will take different values for the independent parameters $\idparam{(i)}$ where each of them is a random realization of the distribution $\pid = \lenspop(\idparam{(i)}|\lensparam)$.

\begin{figure}[h]
\begin{center}
\includegraphics[width=\columnwidth]{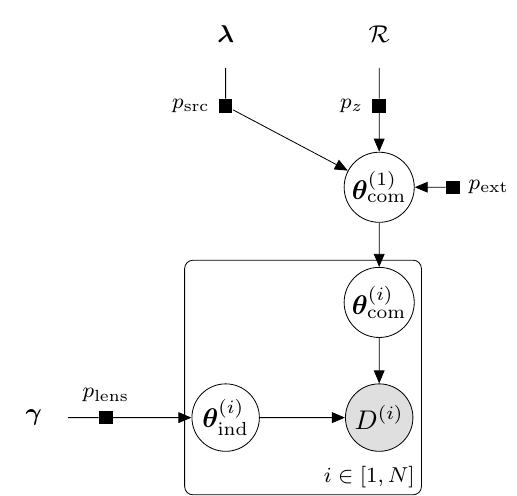}
\end{center}	
\caption{\label{fig:lensed_hyp_data_gen}Data generation process for the $\Ntrig$ observed data under the lensed hypothesis $\lensedhyp$. Each data $\data^{(i)}$ can be described by the event-level parameters $\bm{\eventparam}^{(i)}$, which are partitioned into two disjoint sets: $\comparam{(i)}$ which are assumed to be the same across the $\Ntrig$ signals and $\idparam{(i)}$ which can be different for each signal. Without loss of generality, we assume that $\comparam{(i)} = \comparam{(1)}$ in the graph. The common parameters $\comparam{(1)}$ are one realization of the distribution $\pcom = \srcpop(\comparam{(i)}|\bm{\popparam})p_{z}(z^{(i)}|\mathcal{R}, \lensedhyp) \pext$, while the independent parameters $\idparam{(i)}$ for the $\Ntrig$ signals are $\Ntrig$ realizations of the distribution $\pid = \lenspop(\idparam{(i)}|\lensparam)$.}	
\end{figure}

Figure \ref{fig:lensed_hyp_data_gen} shows a graphical representation of this data generation process. Again one should note that we are not making any inference on the population-level parameters of the GW sources and lenses. Instead we consider them as given in our analysis. Next, we will use our knowledge of the data generation processes under the two hypotheses to construct a statistic that would allow us to evaluate whether some \gls{GW} signals are lensed or not.  

\subsection{Model comparison}
The standard approach to perform a Bayesian model comparison is to compute the posterior odds $\posteriorodds{\lensedhyp}{\notlensedhyp}$, which is defined as\footnote{We will abuse the notation and use $p$ to denote both probability and probability density when the context is clear.}

\begin{equation}
	\posteriorodds{\lensedhyp}{\notlensedhyp} = \frac{p(\lensedhyp|\bm{\data}, \bm{\popparam}, \mathcal{R}, \bm{\lensparam})}{p(\notlensedhyp|\bm{\data}, \bm{\popparam}, \mathcal{R})}.
\end{equation}
Note that for both models we fix the population-level parameters and the merger rate density, and that they are identical for both the lensed and the not-lensed hypothesis. Therefore we will not write them out explicitly when there is no ambiguity. Using Bayes' theorem, we can easily re-write the posterior odds into a product of two terms, namely the Bayes factor and the prior odds as
\begin{equation}
\label{eq:Odds_ratio}
	\posteriorodds{\lensedhyp}{\notlensedhyp} = \underbrace{\frac{p(\bm{\data}|\lensedhyp)}{p(\bm{\data}|\notlensedhyp)}}_{\text{Bayes factor}\; \bayesfactor{\lensedhyp}{\notlensedhyp}} \times  \underbrace{\frac{p(\lensedhyp)}{p(\notlensedhyp)}}_{\text{Prior odds}\; \priorodds{\lensedhyp}{\notlensedhyp}}.
\end{equation}
We first focus on getting an expression for evaluating the Bayes factor $\bayesfactor{\lensedhyp}{\notlensedhyp}$ from the set of $\Ntrig$ observed data. And later we will discuss the evaluation of the prior odds $\priorodds{\lensedhyp}{\notlensedhyp}$.

\subsection{The Bayes factor $\bayesfactor{\lensedhyp}{\notlensedhyp}$}
The Bayes factor $\bayesfactor{\lensedhyp}{\notlensedhyp}$, defined as
\begin{equation}
	\bayesfactor{\lensedhyp}{\notlensedhyp} = \frac{p(\bm{\data}|\lensedhyp)}{p(\bm{\data}|\notlensedhyp)},	
\end{equation}
is a ratio of the normalized probability densities of observing the data set $\bm{\data}$ assuming the two hypotheses under consideration. In Appendix~\ref{app:Full_derivation} we give the full derivation for the expressions evaluating the normalized probability densities of observing the data set $\bm{\data}$ under each of the hypotheses. Here we will outline the derivation. The core idea is to use the graphs that describe the data generation processes for the two hypotheses in Figs.~\ref{fig:not_lensed_hyp_data_gen} and \ref{fig:lensed_hyp_data_gen} to write down the desired expressions for the probability densities, and that the likelihood functions (which are the probability densities viewed as functions of the event-level parameters) can be factorized under both the hypotheses.

For the not-lensed hypothesis, since the $\Ntrig$ signals are independent, we have

\begin{equation}
\label{eq:Factorized_evidence}
	p(\bm{\data}|\notlensedhyp) = \prod_{i=1}^{\Ntrig} p(\data^{(i)}|\notlensedhyp).
\end{equation}
Combining this with the data generation process described in Fig.~\ref{fig:not_lensed_hyp_data_gen} we have

\begin{equation}
	p(\bm{\data}|\notlensedhyp) \propto \prod_{i=1}^{\Ntrig} \int \diff \bm{\eventparam}^{(i)} \; p(\data^{(i)}|\bm{\eventparam}^{(i)}, \notlensedhyp) \ppop(\bm{\eventparam}^{(i)}),
\end{equation}
where the expression on the right-hand side is also known as the (unnormalized) marginal likelihood under the not-lensed hypothesis.
Note that we need to make sure that the probability density $p(\data^{(i)}|\notlensedhyp)$ is normalized over all observable\footnote{Here the term \emph{observable data} means that the signals in the data would pass some detection criteria that one imposes, for example the signals would need to have a signal-to-noise ratio above some threshold, such that they would be identified as \gls{GW} events.} data, accounting for selection effects \cite{2019MNRAS.486.1086M}. This can be done by evaluating the proper normalization constant $\alpha$, where

\begin{equation}
	\alpha = \int_{\text{all obs. data}} \diff \data^{(i)} \; p(\data^{(i)}|\notlensedhyp).
\end{equation}
Therefore the expression for the normalized $p(\bm{\data}|\notlensedhyp)$ is given by

\begin{equation}
\label{eq:Evidence_under_not_lensed_hyp}
	p(\bm{\data}|\notlensedhyp) = \frac{1}{\alpha^{\Ntrig}} \prod_{i=1}^{\Ntrig} \int \diff \bm{\eventparam}^{(i)} \; p(\data^{(i)}|\bm{\eventparam}^{(i)}, \notlensedhyp) \ppop(\bm{\eventparam}^{(i)}).
\end{equation}

As for the lensed hypothesis, unfortunately the probability density $p(\bm{\data}|\lensedhyp)$ cannot be factorized like Eq.~\eqref{eq:Factorized_evidence}. However, the likelihood functions can still be factorized if we assume that the noise realizations for the $\Ntrig$ events are independent and that a signal is deterministic given a set of parameters $\bm{\theta}$ that describe the waveform. Marginalizing the joint likelihood function with parameters according to Fig.~\ref{fig:lensed_hyp_data_gen}, we have
\begin{equation}
\begin{aligned}
	p(\bm{\data}|\lensedhyp) & \propto \int \diff\comparam{(1)} \; \diff\idparam{(1)} \; \cdots \diff\idparam{(\Ntrig)} \\
	& \times \left[ \prod_{j=1}^{\Ntrig} p(\data^{(j)}|\idparam{(j)},\comparam{(1)}) \right] \\
	& \times \pid(\idparam{(1)},...,\idparam{(\Ntrig)}) \pcom(\comparam{(1)}), \\
	&
\end{aligned}
\end{equation}
where the expression on the right-hand side is known as the (unnormalized) marginal likelihood under the lensed hypothesis.
Again, we will need to compute the normalized probability density $p(\bm{\data}|\lensedhyp)$ in order to compute a meaningful Bayes factor, and take selection effects into account. The proper normalization constant $\beta$ in this case, is given by

\begin{equation}
\begin{aligned}
	\beta & \propto \int_{\text{all obs. data set}} \diff \data^{(1)}\cdots\diff\data^{(\Ntrig)} \int \diff\comparam{(1)} \; \diff\idparam{(1)} \; \cdots \diff\idparam{(\Ntrig)} \\
	& \times \left[ \prod_{j=1}^{\Ntrig} p(\data^{(j)}|\idparam{(j)},\comparam{(1)}) \right] \\
	& \times \pid(\idparam{(1)},...,\idparam{(\Ntrig)}) \pcom(\comparam{(1)}).
\end{aligned}
\end{equation}
Therefore, the expression for the normalized $p(\bm{\data}|\lensedhyp)$ is given by
\begin{equation}
\label{eq:Evidence_under_lensed_hyp}
\begin{aligned}
	p(\bm{\data}|\lensedhyp) & =  \frac{1}{\beta} \int \diff\comparam{(1)} \; \diff\idparam{(1)} \; \cdots \diff\idparam{(\Ntrig)} \\
	& \times \left[ \prod_{j=1}^{\Ntrig} p(\data^{(j)}|\idparam{(j)},\comparam{(1)}) \right] \\
	& \times \pid(\idparam{(1)},...,\idparam{(\Ntrig)}) \pcom(\comparam{(1)}). \\
	&
\end{aligned}
\end{equation}

Finally, we have the expression that we can use to evaluate the Bayes factor for the lensed hypothesis versus the not-lensed hypothesis, namely
\begin{widetext}
\begin{equation}
\label{eq:bayes_factor_final_expr}
	\bayesfactor{\lensedhyp}{\notlensedhyp} = \frac{\alpha(\bm{\popparam}, \mathcal{R})^{\Ntrig}}{\beta(\bm{\popparam}, \mathcal{R}, \bm{\lensparam})} \underbrace{\frac{\int \diff\comparam{(1)} \; \diff\idparam{(1)} \; \cdots \diff\idparam{(\Ntrig)} \; \left[ \prod_{j=1}^{\Ntrig} p(\data^{(j)}|\idparam{(j)},\comparam{(1)}) \right] \pid(\idparam{(1)},...,\idparam{(\Ntrig)}|\bm{\lensparam}) \pcom(\comparam{(1)}|\bm{\popparam}, \mathcal{R})}{ \prod_{i=1}^{\Ntrig} \int \diff \bm{\eventparam}^{(i)} \; p(\data^{(i)}|\bm{\eventparam}^{(i)}, \notlensedhyp) \ppop(\bm{\eventparam}^{(i)}|\bm{\popparam}, \mathcal{R}) }}_{\text{coherence ratio} \; \cohratio}.
\end{equation}
\end{widetext}

One can interpret the second factor in Eq.~\eqref{eq:bayes_factor_final_expr}, which is the ratio of unnormalized marginal likelihoods under the two hypotheses, as a measurement of how well the data set $\bm{\data}$ of $\Ntrig$ signals can be jointly fit by a set of common parameters versus $\Ntrig$ sets of independent parameters, which we call it the coherence ratio $\mathcal{C}$ to differentiate it with the Bayes factor. While a negative log-coherence ratio means that the lensed hypothesis, that is setting some of the parameters to be the same across events, fails to fit the $\Ntrig$ signals jointly, a positive log coherence ratio however does not mean that the $\Ntrig$ signals are lensed. This is the Occam's razor at play. Assuming that the lensed hypothesis and the not-lensed hypothesis fit the data set $\bm{\data}$ equally well, the lensed hypothesis will be favored by the Bayesian model selection framework because it has fewer free parameters, and hence a smaller prior volume. For \gls{GW} signals from high-mass \gls{BBH} mergers, this issue will be more apparent as they produce shorter signals detectable in the interferometers, and we usually make less precise measurements of the masses for these high-mass systems \cite{Ghosh:2015jra}. This is partially alleviated by incorporating the population information that they are rarer compared to lighter systems.
It also brings out an important point that the Bayes factor, or generally any probabilistic statement, that some \gls{GW} signals are strongly lensed depends on the source population one is considering.

We can think of the factor $\beta(\bm{\popparam}, \mathcal{R}, \bm{\lensparam})/\alpha(\bm{\popparam}, \mathcal{R})^{\Ntrig}$ in Eq.~\eqref{eq:bayes_factor_final_expr} as a population-averaged scale of the coherence ratio accounting for selection effects, which affect the two hypotheses differently. If the coherence ratio is greater than the population typical value for $\beta/\alpha^{\Ntrig}$, then the Bayes factor will indicate that the lensed hypothesis is favored by the observed data. In fact, the normalization constant under the not-lensed hypothesis $\alpha$ can be interpreted as the detectable fraction of sources \cite{2019MNRAS.486.1086M}. Similarly, we can interpret the normalization constant under the lensed hypothesis $\beta$ as the fraction of sources that would produce $\Ntrig$ detectable lensed signals. We expect that the order of magnitude for $\beta$ would be similar to that for $\alpha$. Therefore, essentially selection effects penalize the lensed hypothesis by a factor of roughly $\alpha^{\Ntrig - 1}$, counteracting the Occam's razor. %

\subsection{The prior odds $\priorodds{\lensedhyp}{\notlensedhyp}$}
The Bayes factor we derived above in Eq.~\eqref{eq:bayes_factor_final_expr} only compares the coherence of the data set with each hypothesis, but not the probability in which each hypothesis would occur. We know empirically, that strong lensing causing at least $\Ntrig$ images occurs less frequently than observing $\Ntrig$ independent \gls{GW} events with each coming from a different source. We can incorporate our knowledge about the rate in the form of prior odds $\priorodds{\lensedhyp}{\notlensedhyp}$, which is defined as
\begin{equation}
\label{eq:prior_odds}
	\priorodds{\lensedhyp}{\notlensedhyp} = \frac{p(\lensedhyp)}{p(\notlensedhyp)}.
\end{equation}
We can then compute the posterior odds $\posteriorodds{\lensedhyp}{\notlensedhyp}$ using Eq.~\eqref{eq:Odds_ratio} from the Bayes factor in Eq.~\eqref{eq:bayes_factor_final_expr} and the prior odds in Eq.~\eqref{eq:prior_odds}.

One can assign the prior odds simply as the ratio of the rate of observing $\Ntrig$ lensed images from a single source over the rate of observing $N$ GW signals coming from $N$ independent sources. Obtaining this will require detailed modeling of GW sources and lenses. In particular these numbers should be computed under the the chosen source and lens population models for an analysis. However, one can argue that for most of the population models commonly used by the astrophysics community, the prior odds is very small with the current sensitivities of GW detectors, about $\priorodds{\lensedhyp}{\notlensedhyp} \approx 10^{-2} - 10^{-4}$ \cite{Ng:2017yiu, Oguri2018, Li:2018prc, Buscicchio:2020cij, Mukherjee:2020tvr}.

\subsection{Marginalization over redshift}
\label{subsec:Marginalization_over_redshift}
With the expression for $p(\bm{\data}|\lensedhyp)$ under the lensed hypothesis in Eq.~\eqref{eq:Evidence_under_lensed_hyp}, one can estimate the integral using a stochastic sampling algorithm such as nested sampling \cite{2004AIPC..735..395S} by sampling over $\left\{ \comparam{(1)}, \idparam{(1)}, \cdots, \idparam{(\Ntrig)} \right\}$ with a prior $\pid(\idparam{(1)},...,\idparam{(\Ntrig)}) \pcom(\comparam{(1)})$ and a \emph{joint} likelihood $\prod_{j=1}^{\Ntrig} p(\data^{(j)}|\idparam{(j)},\comparam{(1)})$. However, a direct sampling will be inefficient because of the degeneracy between the absolute magnification and the luminosity distance, and hence the redshift of the source. Under the not-lensed hypothesis, we can infer the source redshift since we can infer the luminosity distance of the source $d_{\rm L}^{\rm src}$, and by assuming a particular cosmology we can compute the redshift $z^{\rm src} = z(d_{\rm L}^{\rm src})$ from the luminosity distance. Under the lensed hypothesis, each image will be, in general, magnified by a different factor. In fact, we can only measure the apparent luminosity distance for each image as in Eq.~\eqref{eq:Apparent_luminosity_distance}. Therefore, we will not be able to infer the absolute magnification for each image and the source redshift at the same time. For example, a signal with a said redshift of $z \approx 0.363$ and an absolute magnification of $\mu=4$ would have the same apparent luminosity distance of $1$ Gpc as a signal with a redshift of $z \approx 0.780$ and an absolute magnification of $\mu=25$.

In order to explore the degenerate parameter space more efficiently, we can marginalize over the source redshift separately. In fact, the source redshift $z$ stands out from the rest of the parameters. This is because with a given redshift, one can figure out the prior distribution of the apparent luminosity distance $d_{\rm L}^{(i)}$ given the prior distribution of the absolute magnification $p(\mu^{(i)})$ by
\begin{equation}
\begin{aligned}
	p(d_{\rm L}^{(i)}) & = p\left(\left. \mu^{(i)} = \left(\frac{d_{\rm L}^{\rm src}(z)}{d_{\rm L}^{(i)}}\right)^2 \right| z\right) \; \bigg\lvert \frac{\partial \mu^{(i)}}{\partial d_{\rm L}^{(i)}} \bigg\rvert \\
	& = \frac{2\mu^{(i)}}{d_{\rm L}^{(i)}} p\left(\left. \mu^{(i)} = \left(\frac{d_{\rm L}^{\rm src}(z)}{d_{\rm L}^{(i)}}\right)^2 \right| z\right),
\end{aligned}
\end{equation}
and similarly for the prior distribution of the redshifted/detector-frame masses given the distribution of source-frame masses and the redshift as
\begin{equation}
\begin{aligned}
	p(m_{1,2}^{\rm {det}}) & = p(m_{1,2}^{\rm {src}}=\frac{m_{1,2}^{\rm {det}}}{1+z}|z) \bigg\lvert \frac{\partial m_{1,2}^{\rm {src}}}{\partial m_{1,2}^{\rm {det}}} \bigg\rvert \\
	& = \left( \frac{1}{1+z} \right)^{2} p(m_{1,2}^{\rm {src}}=\frac{m_{1,2}^{\rm {det}}}{1+z}|z).
\end{aligned}
\end{equation}
Therefore, we can rewrite Eq.~\eqref{eq:Evidence_under_lensed_hyp} as a 1D integral over the redshift as
\begin{equation}
\label{eq:Marginalization_over_redshift}
	p(\bm{\data}|\lensedhyp) \propto \int \diff z \; \mathcal{L}_{\rm marg}(z) \; p_{z}(z|\lensedhyp),
\end{equation}
where $\mathcal{L}_{\rm marg}(z)$ is given by
\begin{equation}
\begin{aligned}
	\mathcal{L}_{\rm marg}(z) & = \int \diff\comparam{(1)} \setminus \left\{ z \right\} \; \diff\idparam{(1)} \; \cdots \diff\idparam{(\Ntrig)} \\
	& \times \left[ \prod_{j=1}^{\Ntrig} p(\data^{(j)}|\idparam{(j)},\comparam{(1)} \setminus \left\{ z \right\}) \right] \\
	& \times \pid(\idparam{(1)},...,\idparam{(\Ntrig)}) \pcom(\comparam{(1)} \setminus \left\{ z \right\}).
\end{aligned}
\end{equation}
The marginalized likelihood $\mathcal{L}_{\rm marg}(z)$, which is a function of $z$ only, can be obtained via the conventional Monte Carlo methods (such as Markov Chain Monte Carlo method and nested sampling) by sampling over redshifted/detector-frame parameters without the redshift. This will alleviate the degeneracy problem, as well as open up the possibility of computing Eq.~\eqref{eq:Evidence_under_lensed_hyp} by reusing computations done with the not-lensed hypothesis assumed, without re-exploring the joint parameter space. It also lends itself to the interpretation of treating the redshift as a hyperparameter of a subpopulation of signals sharing the same intrinsic parameters (and some of the extrinsic parameters).

\begin{figure}[h]
	\includegraphics[width=\columnwidth]{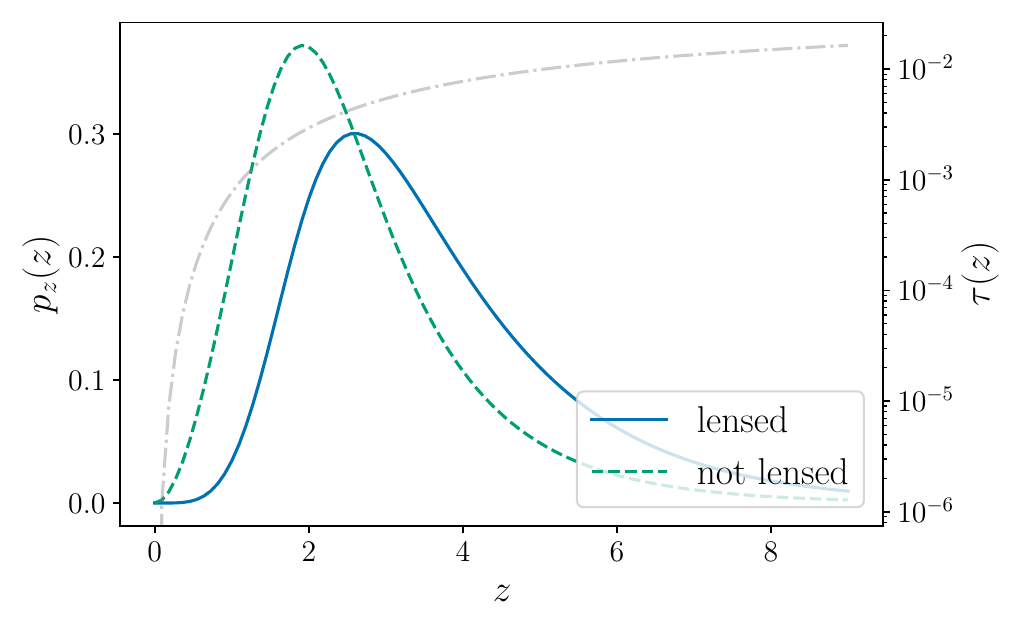}
	\caption{\label{fig:pz}The probability densities $p_{z}(z)$ of the source redshift $z$ under the lensed and not-lensed hypothesis. The gray dotted line shows the optical depth $\tau(z)$. As the optical depth increases with the redshift, the peak of the density $p_{z}$ under the lensed hypothesis shifts to a higher value of $z \sim 3$ compared to the density under the not-lensed hypothesis.}
\end{figure}

Given a merger rate density $\mathcal{R}(z) \equiv \diff N_{\rm src}/\left(\diff V_{\rm c}\diff t\right)$, which is the number density of mergers per comoving volume $V_{\rm c}$ per unit time $t$ in the source frame, one can compute the probability density of the source redshift $z$ as
\begin{equation}
	p(z) \propto \frac{\diff V_{\rm c}}{\diff z} \frac{1}{1+z} \mathcal{R}(z).
\end{equation}
Using the product rule, we can write down the prior distribution for the redshift $z$ under the lensed hypothesis as
\begin{equation}
\label{eq:pz_lensed}
\begin{aligned}
	p_{z}(z|\lensedhyp)	& = \frac{1}{C} \frac{p(\lensedhyp|z) p(z)}{p(\lensedhyp)} \\
	& = \frac{1}{C} \frac{\tau(z)p(z)}{p(\lensedhyp)},
\end{aligned}
\end{equation}
where $C$ is the normalization constant, and $\tau(z) \equiv p(\lensedhyp|z)$ is the optical depth of strong lensing at redshift $z$.
Similarly, under the not-lensed hypothesis, the prior distribution for the redshift $p_{z}(z|\notlensedhyp)$ is given by
\begin{equation}
\label{eq:pz_notlensed}
\begin{aligned}
	p_{z}(z|\notlensedhyp)	& = \frac{1}{C^{'}} \frac{p(\notlensedhyp|z) p(z)}{p(\notlensedhyp)} \\
	& = \frac{1}{C^{'}} \frac{\left[ 1 - \tau(z) \right] p(z)}{p(\notlensedhyp)},
\end{aligned}
\end{equation}
where the normalization constant $C^{'}$ is defined accordingly. Figure \ref{fig:pz} shows the prior distribution of redshift $z$ under the lensed (solid blue line) and not-lensed hypothesis (dashed green line), using the optical depth model in Ref.~\cite{Hannuksela:2019kle} and a merger rate density tracking the star formation rate in \cite{Belczynski:2016ieo, Oguri2018}. The peak of the prior distribution under the lensed hypothesis shifts to a higher value of $z \sim 3$ compared to that under the not-lensed hypothesis, which peaks at roughly $z \sim 2$ because of the optical depth (gray dash-dotted line) being higher at higher redshifts.

As a by-product of evaluating Eq.~\eqref{eq:Marginalization_over_redshift}, we also get a set of posterior samples of $z$, which are distributed according to
\begin{equation}
	p(z|\bm{\data}, \lensedhyp) = \frac{\mathcal{L}_{\rm marg}(z) p_z(z|\lensedhyp)}{\int \diff z \; \mathcal{L}_{\rm marg}(z) \; p_{z}(z|\lensedhyp)}.
\end{equation}
In the next subsection, we describe how to reconstruct the true (but degenerate) source parameters using Gibbs sampling.

\subsection{Inferring source parameters using Gibbs sampling}
\label{subsec:Gibbs_sampling}
Ultimately we want a set of joint posterior samples $\left\{ z, \bm{\bm{\eventparam}} \right\}$ describing the source of the observed lensed signals. As a by-product of the marginalization over the redshift calculation using nested sampling, we obtain a set of posterior samples of the redshift $z \sim p(z|\bm{\data}, \lensedhyp)$ marginalized over the parameters $\bm{\bm{\eventparam}}$. Using Gibbs sampling, we can obtain the desired joint posterior samples from samples drawn from the conditional probability distributions $p(z|\bm{\data}, \lensedhyp)$ from the marginalization step and $p(\bm{\bm{\eventparam}}|z,\bm{\data}, \lensedhyp)$ from the inference step. This is because
\begin{equation}
	p(z, \bm{\bm{\eventparam}}|\bm{\data}, \lensedhyp) \propto p(\bm{\bm{\eventparam}}|z,\bm{\data}, \lensedhyp)p(z|\bm{\data}, \lensedhyp).	
\end{equation}

Algorithm \ref{Algorithm:GibbsSampling} outlines the (collapsed and blocked) Gibbs sampling algorithm \cite{10.2307/2290921} that allows us to reconstruct the joint posterior samples. This variant of Gibbs sampling can be easily parallelized since each iteration is independent. For each set of joint $\left\{ z, \bm{\eventparam} \right\}$ samples, since $\bm{\eventparam}$ are redshifted parameters (such as the redshifted component masses) and $z$ is the source redshift, one can compute the true source parameters such as the source masses easily. In Sec.~\ref{sec:BBH_pair_lensing}, we demonstrate the framework and tools we developed with simulated lensed \gls{GW} signals from \gls{BBH} mergers.

\begin{algorithm}[H]
\caption{Gibbs sampling}\label{Algorithm:GibbsSampling}
\begin{algorithmic}[1]
\Procedure{Sample}{$N_{s}$}
	\State $i \gets 1$
	\State $\bm{\bm{\eventparam}}_{\rm true} \gets [\;]$
	\While{$i \le N_{s}$}
		\State $z_{\rm drawn} \gets$ a random draw from the samples $z$ from the marginalization step
		\State compute the ln weight $\ln w_j$ for each of the samples $\bm{\bm{\eventparam}}_{j}$ from the inference step
		\State $\bm{\bm{\eventparam}}_{\rm drawn} \gets$ a random draw from $\left\{ \bm{\bm{\eventparam}}_j \right\}$ with weight $w_{j}$ using rejection sampling assuming that the true source redshift is $z_{\rm drawn}$
		\State append $\left\{ z_{\rm drawn}, \bm{\bm{\eventparam}}_{\rm drawn} \right\}$ to $\bm{\bm{\eventparam}}_{\rm true}$
		\State $i \gets i + 1$
	\EndWhile
	\State return $\bm{\bm{\eventparam}}_{\rm true}$
\EndProcedure
\end{algorithmic}
\end{algorithm}
 
\section{Strong lensing of gravitational waves from a binary black hole merger: observing a pair of lensed signals}
\label{sec:BBH_pair_lensing}
Now that we have developed the statistical framework in a general setting, here we want to apply the framework to analyze two particular cases and discuss the technical subtleties involved, namely for the case of strong lensing of a \gls{GW} signal from a \gls{BBH} merger with a pair of lensed images (i.e. $\Ntrig = 2$) observed, and with only one image (i.e., $\Ntrig = 1$) observed. In this section, we focus on the former case first.

\subsection{Under the not-lensed hypothesis}
\begin{figure}[h!]
\begin{center}
\includegraphics[width=\columnwidth]{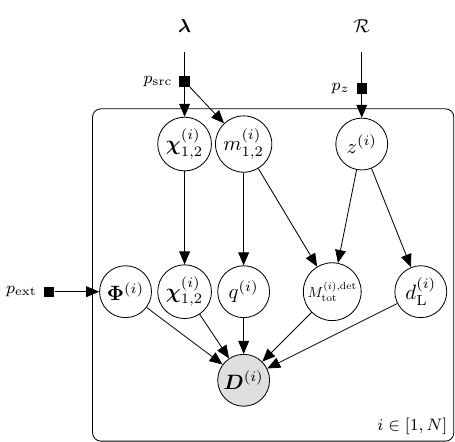}
\end{center}	
\caption{Data generation process for the $\Ntrig$ observed data under the not-lensed hypothesis $\notlensedhyp$. This is similar to Fig.~\ref{fig:not_lensed_hyp_data_gen} but with the event-level parameters $\bm{\eventparam}^{(i)}$ written out explicitly. Here we use $\bm{\Phi}$ to denote the set of extrinsic parameters $\left\{ \alpha, \delta, \psi, \iota, \phi_{c} \right\}$ that are distributed according to the distribution $p_{\rm ext}$.}	
\label{fig:not_lensed_hyp_data_gen_BBH}
\end{figure}

Suppose we write the event-level parameters for each of the BBH mergers under the not-lensed hypothesis as
\begin{equation}
\label{eq:BBH_merger_parametrization}
\begin{aligned}
\bm{\bm{\eventparam}}^{(i)} & = \{\underbrace{M_{\rm {tot}}^{\text{det}}, q, \bm{\chi}_{1}, \bm{\chi}_{2}}_{\text{intrinsic parameters}}, \underbrace{d_{\rm L}, \alpha, \delta, \psi, \iota, \phi_{\rm c}, t_{\rm c}}_{\text{extrinsic parameters}}\},\\
& %
\end{aligned}
\end{equation}
and these are the parameters that are being sampled over during the inference step.
As derived in Eq.~\eqref{eq:Evidence_under_not_lensed_hyp} with $\Ntrig = 2$, under the not-lensed hypothesis we have
\begin{equation}
\begin{aligned}
	& p(\{\data^{(1)}, \data^{(2)}\}|\notlensedhyp) \\
	& = \frac{1}{\alpha^{2}} p(\data^{(1)}|\notlensedhyp) p(\data^{(2)}|\notlensedhyp).
\end{aligned}
\end{equation}
Figure \ref{fig:not_lensed_hyp_data_gen_BBH} shows a graphical representation of the data generation process under the not-lensed hypothesis for signals from \gls{BBH} mergers using the parametrization in Eq.~\eqref{eq:BBH_merger_parametrization}. Here we use $\bm{\Phi}$ to denote the set of extrinsic parameters $\left\{ \alpha, \delta, \psi, \iota, \phi_{c} \right\}$ that are distributed according to the distribution $p_{\rm ext}$. As for the time of arrival $t_{\rm c}$, we treat it separately and hence it is not shown in Fig.~\ref{fig:not_lensed_hyp_data_gen_BBH}. From matched-filtering pipelines that scan through all the data looking for \gls{GW} triggers, we know roughly the time of arrival for each trigger. Let us write $t_{\rm c}^{(1)} = t_{1} + \delta t_{\rm c}^{(1)}$ and $t_{\rm c}^{(2)} = t_{2} + \delta t_{\rm c}^{(2)}$, where $t_1$ and $t_2$ are the point estimates of the arrival times given by a pipeline for the two triggers respectively. Instead of sampling over $t_{\rm c}^{(1)}$ and $t_{\rm c}^{(2)}$, we sample over $\delta t_{\rm c}^{(1)}$ and $\delta t_{\rm c}^{(2)}$ with a small prior range (typically $\sim 0.2$ s) and $t_1, t_2$ taken to be known. Mathematically, this means
\begin{equation}
\label{eq:Transformation_TOA}
\begin{aligned}
	& p(t_{\rm c}^{(1)}, t_{\rm c}^{(2)}|\notlensedhyp) \diff t_{\rm c}^{(1)} \diff t_{\rm c}^{(2)} \\
	& = p(\delta t_{\rm c}^{(1)}, \delta t_{\rm c}^{(2)}|t_{1}, t_{2}, \notlensedhyp) p(t_1, t_2|\notlensedhyp) \diff \delta t_{\rm c}^{(1)} \diff \delta t_{\rm c}^{(2)}.
\end{aligned}
\end{equation}
Suppose we order the two events by their times of arrival, i.e. $t_{2} > t_{1}$, and define the time delay $\Delta t \equiv (t_2 - t_1) > 0$. After this transformation, there is an extra factor in the prior that accounts for the probability of having two random events separated by a time delay of $\Delta t$ under the not-lensed hypothesis.
If we model the arrival of events by a Poisson process, the prior probability density that any random pair of events having a time delay of $\Delta t$, \emph{given that} there are $N_{\rm obs}$ events during the time interval of $(0, T_{\rm obs}]$, is given by
\begin{equation}
	p(\Delta t|\notlensedhyp) = \frac{2}{T_{\rm obs}} \left( 1 - \frac{\Delta t}{T_{\rm obs}} \right),
\end{equation}
where we give a detailed derivation in Appendix~\ref{app:Time_delay_not_lensed}. This can be considered as the part of the time-delay Bayes factor in Ref.~\cite{Haris:2018vmn} from the not-lensed hypothesis.

Therefore, the full expression for $p(\{\data^{(1)}, \data^{(2)}\}|\notlensedhyp)$ now reads
\begin{equation}
\begin{aligned}
	& p(\{\data^{(1)}, \data^{(2)}\}|\notlensedhyp) \\
	& = \frac{1}{\alpha^2} p(\Delta t|\notlensedhyp) \\
	& \times \prod_{i=1}^{2} \int \diff \{ \underbrace{M_{\rm {tot}}^{\text{det}}, q, \bm{\chi}_{1}, \bm{\chi}_{2}, d_{\rm L}, \alpha, \delta, \psi, \iota, \phi_{\rm c}, \delta t_{\rm c}}_{\bm{\eventparam}^{(i)}} \} \\
	& \; \; p(\data^{(i)}|\bm{\eventparam}^{(i)}) \ppop(\bm{\eventparam}^{(i)}|\notlensedhyp),
\end{aligned}
\end{equation}
where under the not-lensed hypothesis there is a one-to-one mapping between $d_{\rm L}$ and $z$, and hence one will only need to convert Eq.~\eqref{eq:pz_notlensed} by multiplying the proper Jacobian without the need of a separate marginalization of the source redshift.

\subsection{Under the lensed hypothesis}
Under the lensed hypothesis, we write the event-level parameters differently, namely we let the common parameters $\comparam{(i)} = \{ M_{\rm {tot}}^{\text{det}}, q, \bm{\chi}_{1}, \bm{\chi}_{2}, \alpha, \delta, \psi, \iota, \phi_{\rm c} \}$. As for the independent parameters, we write $\idparam{(i)} = \{\delta t_{\rm c}^{(i)}, d_{\rm L}^{(i)}, \Xi^{(i)}\}$, where we perform the same transformation to the time of arrival as in the case under the not-lensed hypothesis, and $\Xi$ denotes the type of an image which can be either \{I, II, III\}.

Each strongly lensed image can be classified into three types (I, II or III), where each image type corresponds to a Morse index of $\left\{ 0, 1, 2 \right\}$ respectively, inducing a different phase shift as shown in Eq.~\eqref{eq:Phasing_effect} to the image because of the interaction of the lensed image with the caustic. One would expect the image that arrives at the Earth first to be of type I since type-I images correspond to local minima of the Fermat time-of-arrival potential. However, the signal that we called the first image in an analysis might not actually be the first image that had arrived the Earth since, for example, the GW detectors might be offline. Various arguments on the type of images one would see can be made if we know the geometry of the gravitational lens but this is not known prior to the analysis. Therefore, to be lens-model-agnostic we assume that the type of the lensed images in a pair to each follow a discrete uniform distribution and are uncorrelated, namely
\begin{equation}
    \lenspop(\Xi^{(1)}, \Xi^{(2)}) = \lenspop(\Xi^{(1)}) \lenspop(\Xi^{(2)}),
\end{equation}
where
\begin{equation}
    \lenspop(\Xi^{(i)}) = 
    \begin{cases}
    1/3 \; \text{ when } \Xi^{(i)} = \text{I} \\
    1/3 \; \text{ when } \Xi^{(i)} = \text{II}\\
    1/3 \; \text{ when } \Xi^{(i)} = \text{III}\\
    \end{cases}.
\end{equation}
That being said, both the assumptions that the image types are uncorrelated and each follows a uniform distribution are not true. If one adopts a particular lens model and ordering of the images, the appropriate joint distribution that encapsulates the correlation should be used instead.

\begin{figure}[h!]
\begin{center}
\includegraphics[width=\columnwidth]{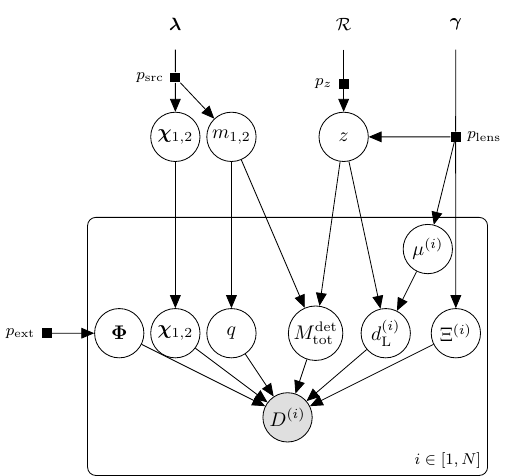}
\end{center}
\caption{Data generation process for the $\Ntrig$ observed data under the lensed hypothesis $\lensedhyp$. This is similar to Fig.~\ref{fig:lensed_hyp_data_gen} but with the common parameters $\comparam{(i)}$ and the independent parameters $\idparam{(i)}$ written out explicitly. Again, we use $\bm{\Phi}$ to denote the set of extrinsic parameters $\left\{ \alpha, \delta, \psi, \iota, \phi_{\rm c} \right\}$ that are distributed according to the distribution $p_{\rm ext}$.}	
\label{fig:lensed_hyp_data_gen_BBH}
\end{figure}

Figure \ref{fig:lensed_hyp_data_gen_BBH} shows a graphical representation of the data generation process under the lensed hypothesis for \gls{BBH} signals. Similar to Fig.~\ref{fig:not_lensed_hyp_data_gen_BBH}, we use $\bm{\Phi}$ to denote the set of extrinsic parameters $\left\{ \alpha, \delta, \psi, \iota, \phi_{\rm c} \right\}$ that are distributed according to the distribution $p_{\rm ext}$, and that we treat the time of arrival $t_{\rm c}$ separately. Unlike the not-lensed case, here we assume that $\comparam{(i)} = \{ M_{\rm {tot}}^{\text{det}}, q, \bm{\chi}_{1}, \bm{\chi}_{2}, \alpha, \delta, \psi, \iota, \phi_{\rm c} \}$ are the same across the signals (hence we dropped the superscript in the graph). Also, even though we sample the apparent luminosity distance for each image, there is no one-to-one mapping between it and the true source redshift since the apparent luminosity distance is also related to the absolute magnification of a lensed image. As discussed in Sec.~\ref{subsec:Marginalization_over_redshift}, we perform the marginalization over the source redshift separately.

For the time of arrival $t_{\rm c}^{(i)}$, we can perform the same transformation as in the case for the not-lensed hypothesis (similar to Eq.~\eqref{eq:Transformation_TOA}), and sample $\delta t_{\rm c}^{(i)}$ that has a much smaller range instead. However, instead of having an analytical expression for the time delay $\Delta t$, there is no analytically tractable expression for the time delay under the lensed hypothesis. That being said, we can obtain it readily from numerical simulations (for example, Ref.~\cite{Haris:2018vmn}). As a result, there is an extra factor of $p(\Delta t|\lensedhyp)$ in the prior that accounts for the probability of having two lensed images separated by a time delay of $\Delta t$.

Therefore, the full expression for $p(\{\data^{(1)}, \data^{(2)}\}|\lensedhyp)$ now reads
\begin{equation}
\begin{aligned}
	& p(\{\data^{(1)}, \data^{(2)}\}|\lensedhyp) \\
	& = \frac{1}{\beta} p(\Delta t|\lensedhyp) \\
	& \times \int \diff z \; p_z(z|\lensedhyp) \Bigg[ \int \diff \{ \underbrace{M_{\rm {tot}}^{\text{det}}, q, \bm{\chi}_{1}, \bm{\chi}_{2}, \alpha, \delta, \psi, \iota, \phi_{\rm c}}_{\comparam{}} \}  \\
	& \int \diff \{ \underbrace{\delta t_{\rm c}^{(1)}, d_{\rm L}^{(1)}, \Xi^{(1)}}_{\idparam{(1)}}, \underbrace{\delta t_{\rm c}^{(2)}, d_{\rm L}^{(2)}, \Xi^{(2)}}_{\idparam{(2)}} \} p(\data^{(1)}|\comparam{}, \idparam{(i)}) \\
	&  p(\data^{(2)}|\comparam{}, \idparam{(2)}) \pid(\idparam{(1)},\idparam{(2)}|\lensedhyp) \pcom(\comparam{}|\lensedhyp) \Bigg],
\end{aligned}
\end{equation}
where the expression enclosed by the square brackets would be identified as $\mathcal{L}_{\rm marg}(z)$ as discussed in Sec~\ref{subsec:Marginalization_over_redshift}.

\subsection{Demonstration}
Here we demonstrate the framework with two examples. In the first example, we injected two \gls{GW} signals with a redshifted total mass $M_{\rm tot}^{\rm det} = 280 M_{\odot}$ into simulated data streams. With this example, we show explicitly how the source population model would change the Bayes factor. In the second example, we injected instead two \gls{GW} signals with a redshifted total mass $M_{\rm tot}^{\rm det} = 60 M_{\odot}$, which corresponds to typical stellar-mass \gls{BBH} systems for the LIGO-Virgo detectors. In both examples, we use the waveform approximant \texttt{IMRPhenomXPHM} \cite{Pratten:2020ceb}, which models both the leading-quadrupole ($\ell = 2$) radiation, as well as some of the nonquadrupole ($\ell > 2$) multipoles. By incorporating the higher-order modes, we show that the image type of each lensed signal can also be inferred.
All the results presented here were computed using the software package \texttt{hanabi}\footnote{\url{https://github.com/ricokaloklo/hanabi}}, which is built upon the package \texttt{bilby} \cite{Ashton:2018jfp} and \texttt{parallel\_bilby} \cite{Smith:2019ucc}. Also, we used the nested sampling algorithm implemented in the package \texttt{dynesty} \cite{Speagle:2019ivv} with the number of live points $\mathtt{nlive} =2000$ and the number of autocorrelation times $\mathtt{nact}=60$ when running the nested sampling algorithm, where the settings are sufficient to give convergent results \cite{Mateu-Lucena:2021siq}. More specifically, the results presented in this paper used the aforementioned packages of version \texttt{bilby==1.0.2} \cite{bilby_2020sep}, \texttt{parallel\_bilby==0.1.5} \cite{pbilby_2020sep}, \texttt{dynesty=1.0.1} \cite{josh_speagle_2019_3461261}, and \texttt{hanabi==0.3.1} \cite{hanabi_github}.

\subsubsection{Example 1: Two lensed signals from apparent intermediate-mass binary black hole mergers}
\label{subsubsec:Example_1}
In this example, we have two lensed \gls{GW} signals injected into two simulated data streams with Gaussian noise recolored to match the \gls{aLIGO} design noise curve \cite{aLIGODesignNoiseCurve}.
Table~\ref{Tab:Injected_value_for_first_example} summarizes some of the waveform parameters for the two signals. The two injected signals, when analyzed on their own, seem to originate from two separate mergers of an intermediate-mass binary black hole system. 

\begin{table}[h]
\caption{\label{Tab:Injected_value_for_first_example}Summary of some of the injection parameters for Example 1 in Sec.~\ref{subsubsec:Example_1}. The two injected signals, when analyzed on their own, seem to originate from two separate mergers of an intermediate-mass binary black hole system. For more detailed definitions of some of the binary parameters, see Ref. \cite{Ashton:2018jfp}.}
\begin{center}
\begin{ruledtabular}
\begin{tabular}{lr}
	Parameter & Value \\
	\colrule
	Redshifted total mass $M_{\rm tot}^{\rm {det}}$ & $280 M_{\odot}$\\
	Mass ratio $q$ & $0.75$ \\
	Redshifted primary mass $m_{1}^{\rm {det}}$ & $160 M_{\odot}$ \\
	Redshifted secondary mass $m_{2}^{\rm {det}}$ & $120 M_{\odot}$ \\
	Dimensionless spin magnitude of the primary $|\bm{\chi}_{1}|$ & 0.3 \\
	Dimensionless spin magnitude of the secondary $|\bm{\chi}_{2}|$ & 0.2 \\
	Tilt angle between the spin vector of the primary & $0.1$ rad \\
	and the orbital angular momentum vector & \\
	Tilt angle between the spin vector of the secondary & $0.2$ rad \\
	and the orbital angular momentum vector & \\
	Azimuthal angle between the two spin vectors & $1.1$ rad \\
	Azimuthal angle of the cone of precession of the orbital & $2.2$ rad \\
	angular momentum about the total angular momentum \\
	Inclination angle between the total angular momentum & $1.04$ rad \\
	and the line of sight & \\
	Right ascension $\alpha$ & $0.2$ rad \\
	Declination $\delta$ & $0.4$ rad \\
	Polarization angle $\psi$ & $0.6$ rad \\
	Phase at coalescence $\phi_{\rm c}$ & $0.8$ rad \\
	Apparent luminosity distance for the & 3.11 Gpc \\
	 first signal $d_{\rm L}^{(1)}$ & \\
	Apparent luminosity distance for the & 3.15 Gpc \\
	second signal $d_{\rm L}^{(2)}$ & \\
	Image type of the first signal & I \\
	Image type of the second signal & II \\
\end{tabular}
\end{ruledtabular}
\end{center}
\end{table}

To demonstrate how using different source population models would change one's interpretation of the two signals, as well as the numerical value of the Bayes factor using our framework, we first use a log-uniform distribution as the population model for the component masses, namely
\begin{equation}
\label{eq:LogUniform_mass}
	\srcpop(m_{1,2}^{\rm {src}}) \propto
	\begin{cases}
		1/m_{1,2}^{\rm {src}} & \text{for } 5 M_{\odot} \leq m_{1,2}^{\rm {src}} \leq 300 M_{\odot} \\
		0 & \text{otherwise}
	\end{cases}.
\end{equation}
For the component spins, we use a distribution that is uniform in the component spin magnitude, and isotropic in the spin orientation.

As for the merger rate density, here we use, for the sake of demonstration, an analytical fit from Ref.~\cite{Oguri2018} that tracks the population synthesis results from Ref.~\cite{Belczynski:2016ieo} for population-I and population-II stars, namely
\begin{equation}
	\mathcal{R}(z) = \frac{6.6 \times 10^{3} \; \exp(1.6z)}{30 + \exp(2.1z)}.
\end{equation}
For the absolute magnification, again for the purpose of demonstration, we use a simple power law distribution that is independent of the time delay, namely
\begin{equation}
\label{eq:Absolute_magnification_prior}
	\lenspop(\mu^{(1)}, \mu^{(2)}|\Delta t) = \lenspop(\mu^{(1)}) \lenspop(\mu^{(2)}),
\end{equation}
with
\begin{equation}
\lenspop(\mu^{(i)}) \propto
	\begin{cases}
	\mu^{-3} & \text{for } \mu \geq 2 \\
	0 & \text{otherwise}	
	\end{cases},
\end{equation}
where it captures the general $\mu^{-3}$ scaling in the high-magnification regime, as well as the requirement that the absolute magnification has to exceed some threshold in order for multiple lensed images to be formed. However, it does not capture the correlation between the magnifications of the lensed images, and the correlation between the magnification and the time delay. For example, the relative magnification tends to unity if the lensed images are highly magnified \cite{Schneider}. In fact, one can derive a poor-man's prior distribution for the relative magnification, if we assume that the absolute magnification for each of the two images follows Eq.~\eqref{eq:Absolute_magnification_prior}, with the form
\begin{equation}
\label{eq:Poorman_prior}
	p(\mu_{\rm rel}) =
	\begin{cases}
		\mu_{\rm rel} & \text{for } \mu_{\rm rel} \leq 1 \\
		\mu_{\rm rel}^{-3} & \text{for } \mu_{\rm rel} > 1 \\
	\end{cases},
\end{equation}
where we give a detailed derivation in Appendix \ref{app:Poorman}. The poor-man's prior distribution for the relative magnification models the correct power law scaling of $\mu_{\rm rel}^{-3}$ when $\mu_{\rm rel} \geq 1$, but predicts the wrong power law scaling when $\mu_{\rm rel} < 1$.

In addition, we use a simple analytical model for the optical depth \cite{1984ApJ...284....1T, Hannuksela:2019kle}, which is the probability of strong lensing at a given redshift, with the form
\begin{equation}
	\tau(z) = F \left( \frac{d_{\rm C}(z)}{d_{\rm H}} \right)^3,
\end{equation}
where $d_{\rm C}(z)$ is the comoving distance at $z$, and $d_{\rm H}$ is the Hubble distance. The empirical constant $F$ is taken to be 0.0017 here \cite{Hannuksela:2019kle}.
A more realistic and detailed model for the merger rate density, the magnification distribution, as well as the optical depth, that impart more astrophysical information to an analysis would certainly help differentiating lensed signals.

\begin{figure*}[ht]
\centering
\includegraphics[width=2\columnwidth]{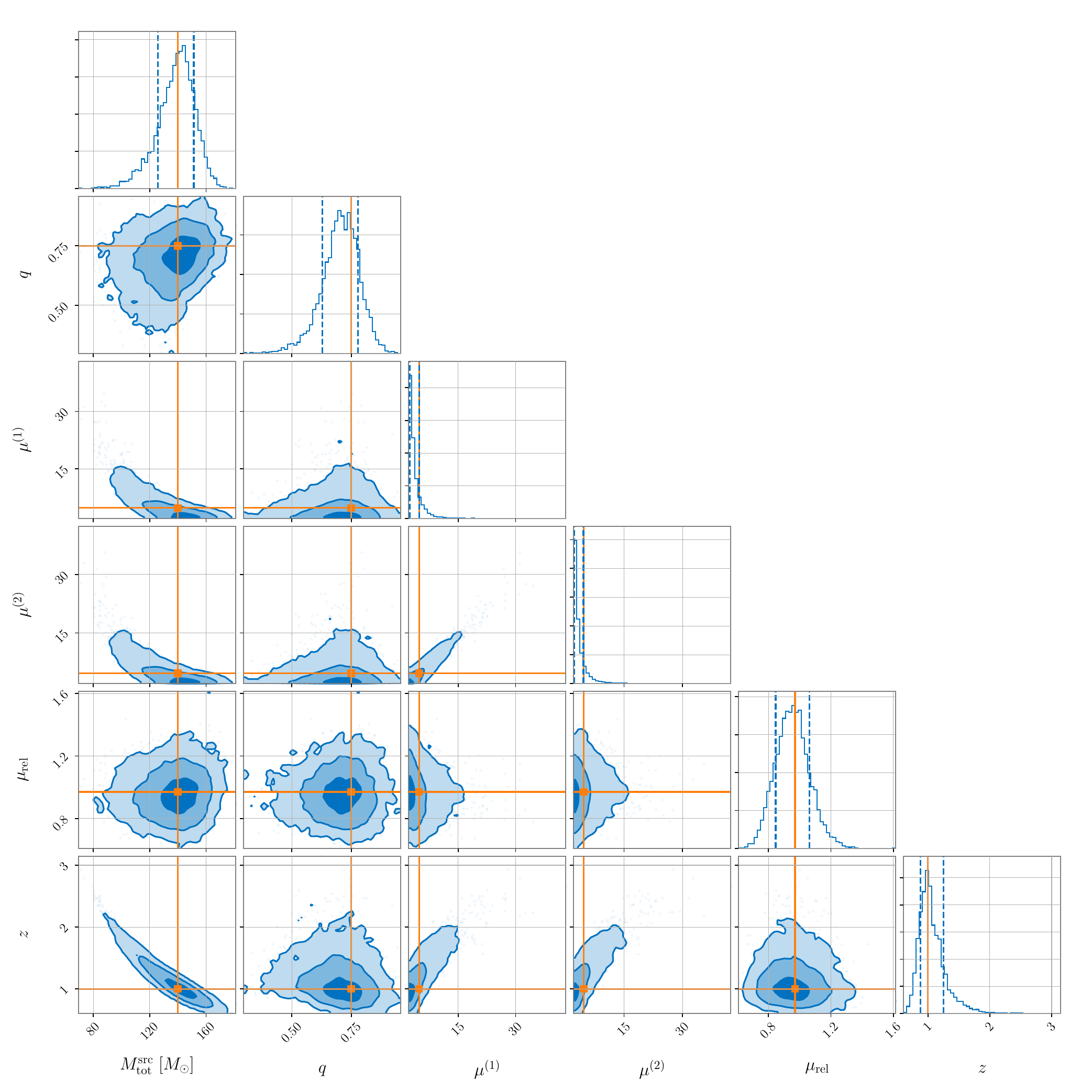}
\caption{\label{fig:Corner_highmass_logunif_mmax300}The 1D and 2D marginalized posterior distributions of $\left\{ M_{\rm tot}^{\rm {src}}, q, \mu^{(1)}, \mu^{(2)}, \mu_{\rm rel}, z \right\}$ for Example 1 (cf. Sec.~\ref{subsubsec:Example_1}) obtained using the algorithm described in Sec.~\ref{subsec:Gibbs_sampling}. The orange solid lines show the correct values for each of the parameters if the redshift $z$ is set to $1$. The plot shows that our two-step hierarchical procedure described in Sec. \ref{subsec:Marginalization_over_redshift} is able to find the correct values describing the signals. From the plot we can also see the various degeneracies between parameters. For example, the degeneracy between the total mass $M_{\rm tot}^{\rm {src}}$ and the redshift $z$, where the blob in the lower left corner corresponds to the redshifted total mass that we do measure. Note that we are able to infer the mass ratio $q$ and the relative magnification $\mu_{\rm ref}$ as they are not degenerate with the redshift.}
\end{figure*}

With this set of population models, we obtained a log-coherence ratio of $\log_{10} \cohratio = 2.7$, and a log-Bayes factor of $\log_{10} \bayesfactor{\lensedhyp}{\notlensedhyp} = 1.1$ without accounting for the time delay.\footnote{This means that we set $p(\Delta t|\lensedhyp)/p(\Delta t|\notlensedhyp) = 1$ such that the coherence ratios and the Bayes factors reported will \emph{not be boosted} by setting the time-delay of the injected signals to be very consistent with the lensed hypothesis, though incorporating the time delay information would be important in actual analyses.} We see that with this set of population models and the detector sensitivity, the selection effects down-weight the pair by a factor of $\approx 40$. Figure \ref{fig:Corner_highmass_logunif_mmax300} shows both the 1D and 2D marginalized posterior distributions for $\left\{ M_{\rm tot}^{\rm {src}}, q, \mu^{(1)}, \mu^{(2)}, \mu_{\rm rel}, z \right\}$ obtained using the algorithm described in Sec.~\ref{subsec:Gibbs_sampling}. The orange solid lines show the correct values for each of the parameters if the redshift $z$ is set to $1$. The plot shows that our two-step hierarchical procedure described in Sec. \ref{subsec:Marginalization_over_redshift} is able to find the correct values describing the signals. From the plot we can also see the various degeneracies between parameters. For example, the degeneracy between the total mass $M_{\rm tot}^{\rm {src}}$ and the redshift $z$, where the blob in the lower left corner of Fig.~\ref{fig:Corner_highmass_logunif_mmax300} corresponds to the redshifted total mass that we do measure. Note that we are able to infer the mass ratio $q$ and the relative magnification $\mu_{\rm ref}$ as they are not degenerate with the redshift.

If we instead use a population model that asserts there are no black holes with mass greater than $60M_{\odot}$, referred as ``Model A'' in Ref.~\cite{LIGOScientific:2018jsj}, namely
\begin{widetext}
\begin{equation}
\label{eq:PopModel_modelA}
\begin{aligned}
    \srcpop(m_1^{\rm {src}}, m_2^{\rm {src}}|\alpha, \beta, m_{\rm min}, m_{\rm max}) = \begin{cases}
    	\frac{1 - \alpha}{m_{\rm max}^{1-\alpha} - m_{\rm min}^{1-\alpha}} (m_1^{\rm {src}})^{-\alpha} \frac{1 + \beta}{(m_{1}^{\rm {src}})^{1+\beta} - m_{\rm min}^{1+\beta}} (m_2^{\rm {src}})^{\beta} & \textrm{if } m_{\rm min} \leq m_2^{\rm {src}} \leq m_1^{\rm {src}} \leq m_{\rm max} \\
    	0 & \textrm{otherwise}
    \end{cases},
\end{aligned}
\end{equation}
\end{widetext}
with $\alpha=1.8$, $\beta=0$, $m_{\rm min}=5\;M_{\odot}$, and $m_{\rm max}=60\;M_{\odot}$\footnote{Note that the numbers we adopted here are slightly different from the reported values in Ref.~\cite{LIGOScientific:2018jsj} because we write the model in the $(m_{1}^{\rm {src}}, m_{2}^{\rm {src}})$ parametrization here instead of the $(m_{1}^{\rm {src}}, q)$ parametrization. The probability density is hence off by a Jacobian of $1/m_1^{\rm {src}}$, which can be easily accounted for by adjusting the value of $\alpha$.}, now both the log-coherence ratio and the log-Bayes factor are infinite, while the log evidence under the lensed hypothesis is finite. This is a ``smoking-gun evidence'' that the two signals are lensed.
This is not surprising because the two signals are impossible under the not-lensed hypothesis with this set of population models. Under the not-lensed hypothesis, we interpret the apparent luminosity distance as the true luminosity distance without any magnification bias, allowing us to infer the redshift directly from the measured luminosity distance. In this case, the redshift that corresponds to the apparent luminosity distance of the first signal is roughly $z \approx 0.53$, meaning that both the primary and secondary mass would be above the $60M_{\odot}$ maximum. This example, though extreme, clearly shows that the Bayes factor, and hence one's interpretation on the origin, of the signals would be sensitive to the population models that one assumes.

\subsubsection{Example 2: Two lensed signals from a stellar-mass binary black hole merger}
\label{subsubsec:Example_2}
In the second example, we also have two lensed \gls{GW} signals injected into two simulated data streams with Gaussian noise. However, this time the two signals have a lower redshifted total mass ($M_{\rm tot}^{\rm {det}} = 60M_{\odot}$). Table \ref{Tab:Injected_value_for_second_example} summarizes some of the waveform parameters. This example serves to represent typical scenarios for second-generation terrestrial GW detectors such as the two Advanced LIGO detectors \cite{TheLIGOScientific:2014jea} and the Advanced Virgo detector \cite{TheVirgo:2014hva} observing stellar-mass \gls{BBH} systems, and demonstrate how would the Bayes factor change with different detector sensitivities. For the population models, we use the same set of models in the last subsection with the ``Model A'' mass model described in Eq.~\eqref{eq:PopModel_modelA}.

\begin{table}[h]
\caption{\label{Tab:Injected_value_for_second_example}Summary of some of the injection parameters for Example 2 in Sec.~\ref{subsubsec:Example_2}. This example serves to represent typical scenarios for second-generation terrestrial GW detectors observing stellar-mass \gls{BBH} systems. Parameters for this injection that are not listed explicitly below are identical to that listed in Table \ref{Tab:Injected_value_for_first_example}.}
\begin{center}
\begin{ruledtabular}
\begin{tabular}{lr}
	Parameter & Value \\
	\colrule
	Redshifted total mass $M_{\rm tot}^{\rm {det}}$ & $60 M_{\odot}$\\
	Mass ratio $q$ & $0.875$ \\
	Redshifted primary mass $m_{1}^{\rm {det}}$ & $32 M_{\odot}$ \\
	Redshifted secondary mass $m_{2}^{\rm {det}}$ & $28 M_{\odot}$ \\
	Apparent luminosity distance for the & 811 Mpc \\
	first signal $d_{\rm L}^{(1)}$ & \\
	Apparent luminosity distance for the & 823 Mpc \\
	second signal $d_{\rm L}^{(2)}$ & \\
\end{tabular}
\end{ruledtabular}
\end{center}
\end{table}

\begin{figure*}[ht]
\centering
\includegraphics[width=2\columnwidth]{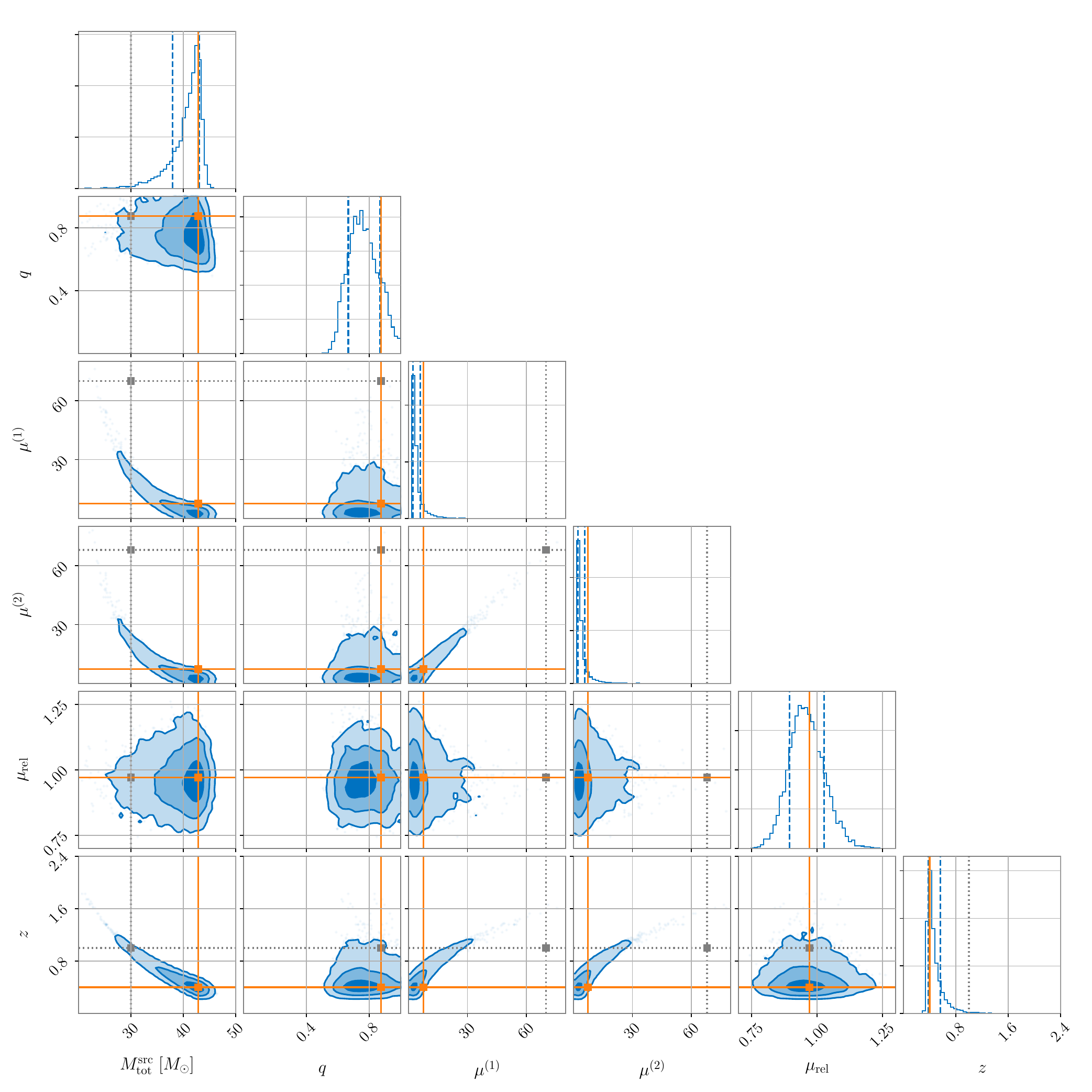}
\caption{\label{fig:Corner_detframe60}The 1D and 2D marginalized posterior distributions of $\left\{ M_{\rm tot}^{\rm {src}}, q, \mu^{(1)}, \mu^{(2)}, \mu_{\rm rel}, z \right\}$ for Example 2 (cf. Sec.~\ref{subsubsec:Example_2}) obtained using the algorithm described in Sec.~\ref{subsec:Gibbs_sampling}. The orange solid lines show the correct values for each of the parameters if the redshift is set to $z=0.4$, while the gray dotted lines show the correct values for the parameters if the redshift is instead set to $z=1$. Note that both the mass ratio $q$ and the relative magnification $\mu_{\rm rel}$ take the same value when different source redshifts are assumed. While we are not able to constrain the source parameters individually because of the degeneracies, we are capable of providing joint constraints for the source parameters by properly incorporating information from both the detected signals and the astrophysical population models assumed. We see that it is less likely for the signals to come from a binary system with a total mass of $M_{\rm tot}^{\rm {src}} = 30\;M_{\odot}$ at a redshift $z = 1$ under the lensed hypothesis because of the large absolute magnifications required are less probable under the lens model we assumed in the analysis.
}
\end{figure*}

Figure~\ref{fig:Corner_detframe60} shows the marginalized 1D and 2D posterior distributions for $\left\{ M_{\rm tot}^{\rm {src}}, q, \mu^{(1)}, \mu^{(2)}, \mu_{\rm rel}, z \right\}$ we recover when the two lensed signals were injected into data streams with simulated Gaussian noise recolored to match the \gls{aLIGO} design sensitivity \cite{aLIGODesignNoiseCurve}. From the plot we see similar degenerate structures between parameters as in Fig.~\ref{fig:Corner_highmass_logunif_mmax300}. To demonstrate the degeneracies more explicitly, we show the correct source parameters for this two signals if we assume the true source redshift is $z=0.4$ (solid orange lines), as well as that if the true redshift is instead $z=1$ (dotted gray lines). Note that both the mass ratio $q$ and the relative magnification $\mu_{\rm rel}$ take the same value when different source redshifts are assumed. While we are not able to constrain the source parameters individually because of the aforementioned degeneracies, we are capable of providing joint constraints for the source parameters by properly incorporating information from both the detected signals and the astrophysical population models assumed. From Fig.~\ref{fig:Corner_detframe60}, we see that it is less likely for the signals to come from a binary system with a total mass of $M_{\rm tot}^{\rm {src}} = 30M_{\odot}$ at a redshift $z = 1$ under the lensed hypothesis because of the large absolute magnifications required are less probable under the lens model we assumed in the analysis.

For this example, we obtained a log-coherence ratio of $\log_{10} \mathcal{C} = 5.2$ and a log-Bayes factor of  $\log_{10} \bayesfactor{\lensedhyp}{\notlensedhyp} = 3.0$ when injecting the signals into simulated Gaussian noise recolored to match the \gls{aLIGO} design sensitivity \cite{aLIGODesignNoiseCurve}. Table \ref{Tab:Selection_functions_different_sensitivities} tabulates the values of $\log_{10} \alpha$, $\log_{10} \beta$, and $\log_{10} \left( \beta/\alpha^2 \right)$ for this particular set of population models under different detector sensitivities, computed using \texttt{pdetclassifier} \cite{Gerosa:2020pgy}. In Appendix \ref{app:Evaluation_of_selection_functions}, we give a detailed description on how one can compute these normalization constants, or selection functions, under the lensed and the not-lensed hypothesis.
\begin{table}[h!]
\caption{\label{Tab:Selection_functions_different_sensitivities}The values of $\log_{10} \alpha(\bm{\popparam}, \mathcal{R})$, $\log_{10} \beta(\bm{\popparam}, \mathcal{R}, \bm{\lensparam})$, and $\log_{10} \left[  \beta(\bm{\popparam}, \mathcal{R}, \bm{\lensparam})/\alpha(\bm{\popparam}, \mathcal{R})^2 \right]$ with different detector sensitivities computed using \texttt{pdetclassifier} \cite{Gerosa:2020pgy} for the population models described in Sec.~\ref{subsubsec:Example_2}.}
\begin{ruledtabular}
\begin{tabular}{lrrr}
	& O1+O2 & O3a & \gls{aLIGO} design \\
	\tableline
	$\log_{10} \alpha$ & $-3.5$ & $-3.1$ & $-2.4$ \\
	$\log_{10} \beta$ & $-4.1$ & $-3.7$ & $-2.5$ \\
	$\log_{10} \left( \beta/\alpha^2 \right)$ & 2.9 & 2.5 & 2.3
\end{tabular}
\end{ruledtabular}
\end{table}
As expected, the values of $\alpha$ and $\beta$ increase as the detector network becomes more sensitive and capable of detecting weaker signals. The difference between the values of $\alpha$ and $\beta$ narrows as the network increases in sensitivity, and that the selection effects penalize the lensed hypothesis to a lesser extent, roughly by a factor of $\sim \alpha$. While we did not perform the same injection test with simulated noise recolored to match the sensitivity during O1+O2 and O3a, we can reasonably expect the log-coherence ratio increases with a more sensitive detector network as we can better measure the waveform parameters to a higher precision. Therefore, the log-coherence ratio, as well as the log-Bayes factor would increase with the detector sensitivity given the same set of lensed signals.

\subsection{Identifying the image types}
\begin{figure*}[ht]
\subfloat[Using \texttt{IMRPhenomXPHM} \cite{Pratten:2020ceb} waveform model. It includes some higher modes $(\ell > 2)$ other than the quadrupole $(\ell = 2)$ radiation.]{\includegraphics{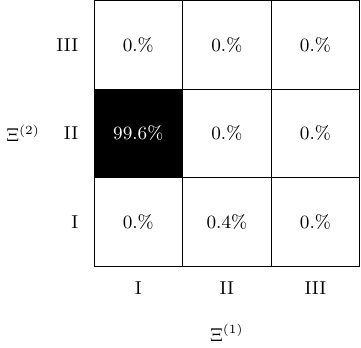}}\;\;\;\;\;
\subfloat[Using \texttt{IMRPhenomXP} \cite{Pratten:2020ceb} waveform model. It only models the quadrupole ($\ell = 2)$ radiation.]{\includegraphics{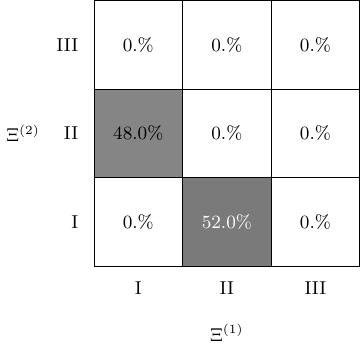}}
\caption{\label{fig:Image_type}The joint posterior probability mass function of the (discrete) image type for the first signal $\Xi^{(1)}$ and that for the second signal $\Xi^{(2)}$ in an injection test. In the test, we injected a type-I signal into the first data stream, and a type-II signal into the second data stream. We see that when there are measurable contributions from higher modes, we are able to pinpoint the type of each lensed image, breaking the degeneracy between the phasing effect from strong lensing and a shift in the orbital phase. This is in line with the findings in Ref.~\cite{Wang:2021kzt}.}
\end{figure*}

When we consider only the dominant $\ell=|m|=2$ modes and a nonprecessing binary system, the phasing effect due to strong lensing reduces to a shift in the observed phase at coalescence (or any reference orbital phase) \cite{Dai:2017huk, Ezquiaga:2020gdt}. For a \gls{GW} signal from the merger of a precessing binary system with a significant contribution from higher-order modes, for example when the system is asymmetric in component masses and/or is inclined with respect to our line of sight, we can break the degeneracy between the phasing effect from strong lensing and the orbital phase. This allows us to identify the image type for each of the lensed signals.
We demonstrate this by injecting signals with an asymmetric mass ratio $q \approx 0.3$ viewing at an angle of roughly $107 \deg$ between the line of sight and the total angular momentum vector into simulated Gaussian noise at \gls{aLIGO} design sensitivity using two different waveform models, \texttt{IMRPhenomXP} and \texttt{IMRPhenomXPHM} \cite{Pratten:2020ceb}. The former approximant, \texttt{IMRPhenomXP}, includes only the quadrupole ($\ell = 2$) radiation from a precessing binary system, while the latter approximant, \texttt{IMRPhenomXPHM}, includes both the quadrupole radiation and some of the higher multipoles ($\ell > 2$) from the precessing system. In both cases, the first injected lensed \gls{GW} signal is of type I, while the second injected signal is of type II. Figure \ref{fig:Image_type} shows the joint probability mass function of the image type inferred for the first signal $\Xi^{(1)}$ and that for the second signal $\Xi^{(2)}$. We see that when there are measurable contributions from higher modes, we are able to pin-point the type of each lensed image from the phasing effect (left panel of Fig.~\ref{fig:Image_type}), breaking the degeneracy between the phasing effect from strong lensing and the shift in the orbital phase. This is in line with the findings reported in Ref.~\cite{Wang:2021kzt}, where one can tell type-II images apart individually for third-generation detectors.

\subsection{Improvement in localizing the source in the sky}
\begin{figure}[h!]
\centerline{
\includegraphics[width=1.1\columnwidth]{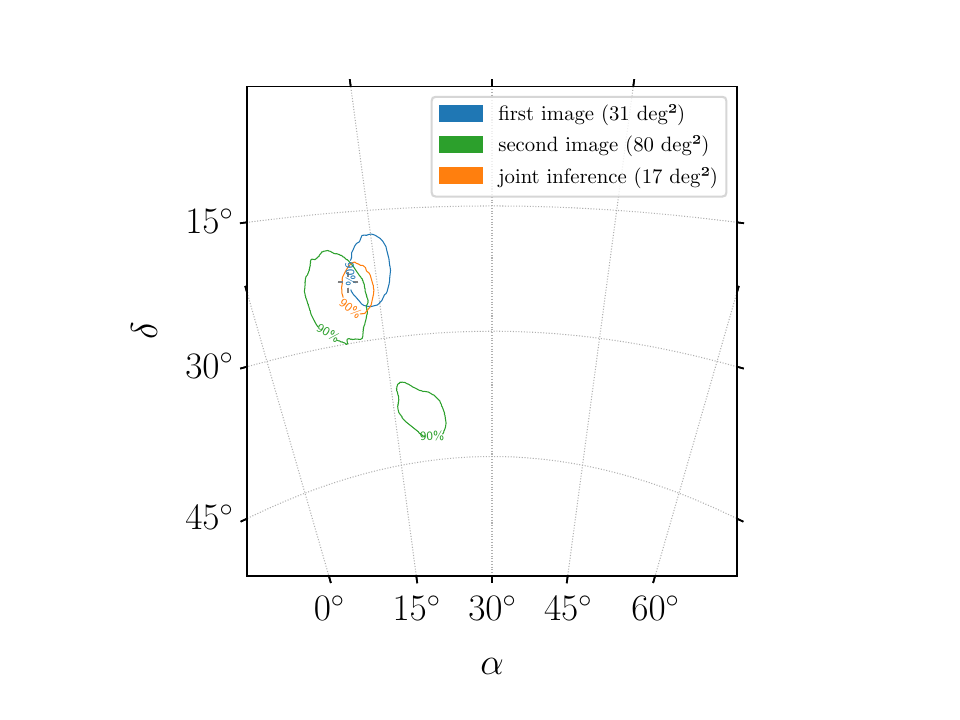}}
\caption{\label{fig:skymaps}The sky localizations when two simulated lensed GW signals are analyzed jointly and when they are analyzed individually. The gray crosshair shows the injected values for the right ascension $\alpha$ and the declination $\delta$. The signals are better localized when analyzed jointly (area of the $90\%$ credible region: $17 \deg^{2}$) compared to the case when they are analyzed individually (area of the $90\%$ credible region: $31 \deg^{2}$ for the brighter image, $80 \deg^2$ for the fainter image) as expected \cite{Seto:2003iw} since we gain information about the shared sky location from two data streams instead of just one. The improvement in sky localization helps identifying the gravitational lens and the source electromagnetically \cite{Seto:2003iw, Hannuksela:2020xor}, and hence cross-validating the claim that the GW signals are indeed strongly lensed.}
\end{figure}

Since we expect the lensed \gls{GW} signals coming from the same source to have approximately identical sky locations, the signals should be better localized when analyzed jointly compared to the case when they are analyzed individually. This is because we gain information about the shared sky location from two data streams instead of just one. 
We demonstrate this using the inference results from Example 1 in Sec.~\ref{subsubsec:Example_1}.
Figure~\ref{fig:skymaps} shows the $90\%$ credible regions of the localization of signals, when analyzed separately (blue and green) and when analyzed jointly (orange). In all cases, the credible regions enclose the true source location (gray crosshair). However, the area of the $90\%$ credible region, a metric for the localization uncertainty, from the joint inference is only $17 \deg^{2}$, which is roughly two times smaller than that when localizing the first image only ($31 \deg^2$) and roughly four times smaller than that when localizing the second, fainter, image only ($80 \deg^2$).

Combining the improved sky localization of the source with the joint constraints of the source parameters (such as the redshift), one will be more informed when trying to locate the gravitational lens and the source electromagnetically (see, for example, Ref.~\cite{Seto:2003iw, Hannuksela:2020xor}). Indeed, if we were able to identify the massive object responsible for the gravitational lensing and observe lensing of electromagnetic waves as well, that can serve as a cross-validation that the \gls{GW} signals that were being analyzed are indeed strongly lensed. The lensed source could be by itself electromagnetically bright, for example a binary neutron star system or a neutron-star-black-hole binary (though it is not expected to see lensed signals coming from these kind of sources with current-generation \gls{GW} detectors as they have lower merger rates relative to \gls{BBH} systems). If the lensed source is in a host which emits electromagnetic radiation, such as in a galaxy, one could search for possible foreground lenses (as the electromagnetic signature coming from the host should also be lensed by the same lens) within the joint localization region that would produce lensed \gls{GW} signals with consistent absolute magnifications and time delays.
 
\section{Strong lensing of gravitational waves from a binary black hole merger: observing only one lensed signal}
\label{sec:BBH_single_lensing}
The statistical framework can also be applied when only one \gls{GW} signal (i.e. $\Ntrig = 1$) is being analyzed at a time. In this case, the expression for the Bayes factor reads
\begin{widetext}
\begin{equation}
\begin{aligned}
	\bayesfactor{\lensedhyp}{\notlensedhyp} = \frac{\alpha(\bm{\popparam}, \mathcal{R})}{\beta(\bm{\popparam}, \mathcal{R}, \bm{\lensparam})} \frac{\int \diff\comparam{(1)} \; \diff\idparam{(1)} \; p(\data^{(1)}|\idparam{(1)},\comparam{(1)}) \pid(\idparam{(1)}|\bm{\lensparam}) \pcom(\comparam{(1)}|\bm{\popparam}, \mathcal{R})}{ \int \diff \bm{\eventparam}^{(1)} \; p(\data^{(1)}|\bm{\eventparam}^{(1)}, \notlensedhyp) \ppop(\bm{\eventparam}^{(1)}|\bm{\popparam}, \mathcal{R}) },
\end{aligned}
\end{equation}
\end{widetext}
where the normalization constant $\beta(\bm{\popparam}, \mathcal{R}, \bm{\lensparam})$ for the case of $\Ntrig = 1$ is defined accordingly. For \gls{BBH} systems, the data generation process described in Fig.~\ref{fig:not_lensed_hyp_data_gen_BBH} for the not-lensed hypothesis, and that in Fig.~\ref{fig:lensed_hyp_data_gen_BBH} for the lensed hypothesis, are also applicable here. Compared with the case of $\Ntrig = 2$ we discussed extensively in Sec.~\ref{sec:BBH_pair_lensing}, the framework is less capable of differentiating a lensed \gls{BBH} signal in the geometric optics limit from a signal that is not lensed. This is because effectively the framework is leveraging only the inconsistency of the signal with the given population models without the help of Occam's razor and selection effects. However, this will not be the case if gravitational lensing leaves distinctive signatures in the observed waveforms, for example when the geometric optics approximation breaks down and the full wave optics treatment is needed \cite{Schneider, Takahashi:2003ix}. The framework can also be easily extended to handle \gls{GW} lensing from BNS systems (see, for example, Ref.~\cite{Pang:2020qow}) and NSBH systems.

Here we demonstrate the statistical framework when $\Ntrig = 1$ with an example where a lensed \gls{BBH} signal with waveform parameters identical to the first image of Example 1 in Sec.~\ref{subsubsec:Example_1} (cf. Table \ref{Tab:Injected_value_for_first_example}) injected into simulated Gaussian noise recolored to the \gls{aLIGO} design noise curve \cite{aLIGODesignNoiseCurve}. As we shall see, the framework is less capable of identifying a lensed \gls{BBH} signal purely from its inconsistency with the population models, unless the models fail to produce such an observed signal (such as Example 1 in Sec.~\ref{subsubsec:Example_1} with a source population model asserting that no black hole can have a mass greater than $60M_{\odot}$). This is very similar to the lensing analysis for GW190521 presented in Ref.~\cite{Abbott:2020mjq}, and the BNS lensing analysis in Ref.~\cite{Pang:2020qow}.

\begin{figure*}[ht]
\centering
\includegraphics[width=1.85\columnwidth]{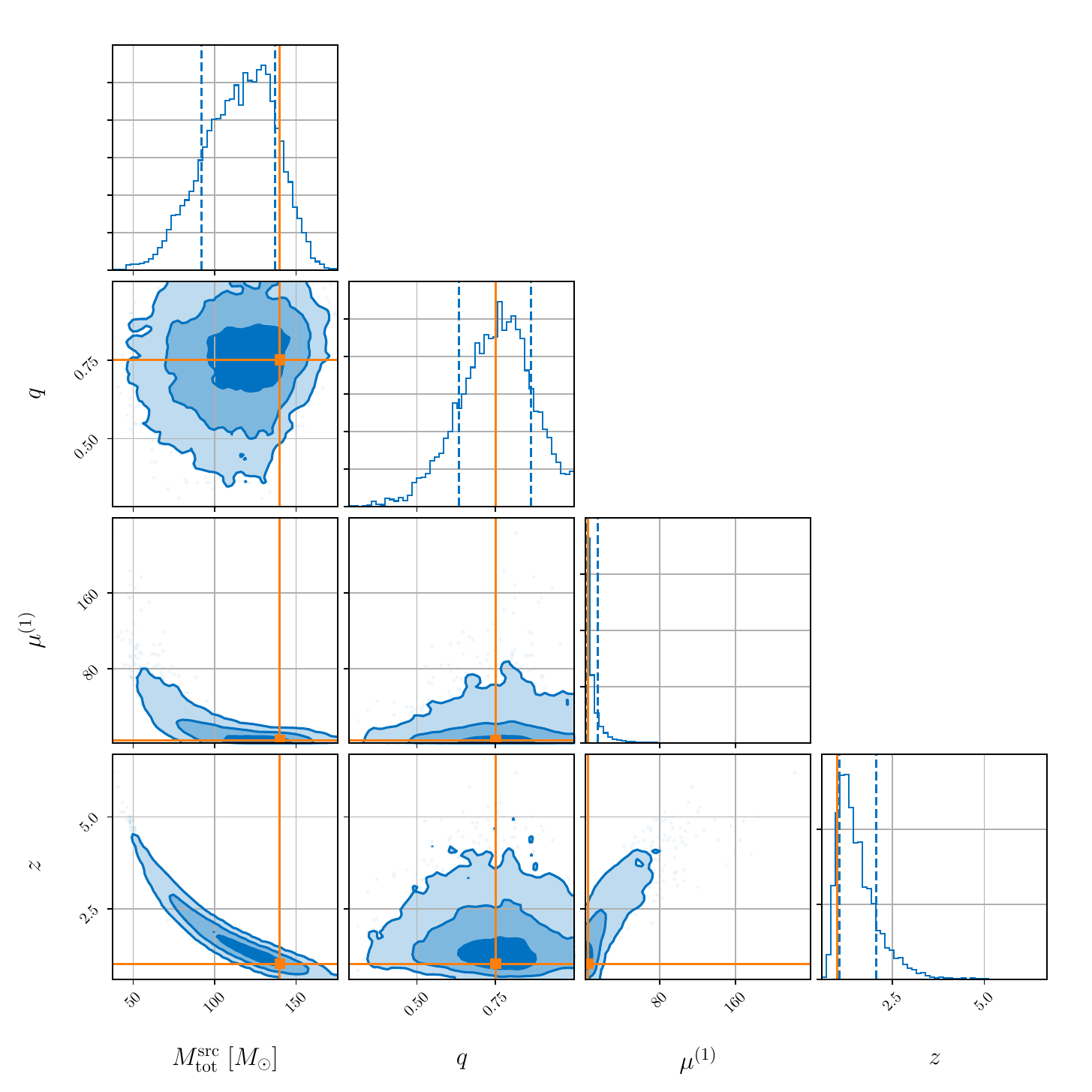}
\caption{\label{fig:Corner_single_highmass_BBH}The 1D and 2D marginalized posterior distributions of $\left\{ M_{\rm tot}^{\rm {src}}, q, \mu^{(1)}, z \right\}$ for the example in Sec.~\ref{sec:BBH_single_lensing} obtained using the algorithm described in Sec.~\ref{subsec:Gibbs_sampling}. The orange solid lines show the correct values for each of the parameters if the redshift $z$ is set to $1$. Similar to the case with $\Ntrig = 2$ in Fig.~\ref{fig:Corner_highmass_logunif_mmax300}, we observe similar degenerate structures such as that between $M_{\rm tot}^{\rm {src}}$ and $z$. However, the uncertainty in the joint $M_{\rm tot}^{\rm {src}}-z$ constraint here when $\Ntrig = 1$ is greater compared with that in Fig.~\ref{fig:Corner_highmass_logunif_mmax300}. This can be attributed to the less constraining measurements of the redshifted masses when there is only one data stream to infer from, instead of two data streams as in the case for Fig.~\ref{fig:Corner_highmass_logunif_mmax300}.}
\end{figure*}

In this example, we first use the log-uniform distribution in Eq.~\eqref{eq:LogUniform_mass} as the population model for the component masses, and use the same models for spin, magnification, optical depth, and merger rate density as in Sec.~\ref{subsubsec:Example_1}. Figure~\ref{fig:Corner_single_highmass_BBH} shows the marginalized 1D and 2D posterior distributions of $\left\{ M_{\rm tot}^{\rm {src}}, q, \mu^{(1)}, z \right\}$, again using the Gibbs sampling algorithm described in Sec.~\ref{subsec:Gibbs_sampling}. The orange solid lines show the correct values for each of the parameters if the redshift $z$ is set to $1$. Similar to the case with $\Ntrig = 2$ in Fig.~\ref{fig:Corner_highmass_logunif_mmax300}, we observe similar degenerate structures such as that between $M_{\rm tot}^{\rm {src}}$ and $z$. However, the uncertainty in the joint $M_{\rm tot}^{\rm {src}}-z$ constraint here when $\Ntrig = 1$ is greater compared to that in Fig.~\ref{fig:Corner_highmass_logunif_mmax300}. This can be attributed to the less constraining measurements of the redshifted masses when there is only one data stream to infer from, instead of two data streams as in the case for Fig.~\ref{fig:Corner_highmass_logunif_mmax300}.

With this set of population models, we obtained a log-coherence ratio of $\log_{10} \mathcal{C} = 0.06$, and a log-Bayes factor of $\log_{10} \bayesfactor{\lensedhyp}{\notlensedhyp} = 0.08$.
The Bayes factor here is not statistically significant and within the statistical uncertainty from using nested sampling. Given the rarity of strong lensing, reflected in the prior odds, the strong lensing hypothesis will easily be dismissed with the small posterior odds. While the above example might give a pessimistic impression for identifying strongly lensed \gls{BBH} signals individually, this does not have to be the case, especially if the signals are of type II, where we might be able to identify them from the distortion to the waveforms due to lensing, for example from the extra Morse phase \cite{Wang:2021kzt}, or other frequency-dependent wave optics effects.

\section{Conclusions and outlook}
\label{sec:conclusions}
In this \this, we present a Bayesian statistical framework for identifying strongly lensed GW signals. By modeling the data generation processes for $\Ntrig$ observed signals assuming that they are lensed images from the same source and that they are simply $\Ntrig$ signals from $\Ntrig$ different sources respectively, we are able to write down an expression for the Bayes factor which quantifies the ratio of the probability densities of observing the $\Ntrig$ signals given the two scenarios and the astrophysical models on the properties of the sources and lenses. Selection effects are accounted for naturally as we normalize the probability densities. Combined with the prior odds, we can properly interpret the resultant posterior odds as the ratio of the probability that the $\Ntrig$ signals are strongly lensed versus not lensed.

In order to compute the marginalization integral for the Bayes factor more efficiently, we present a hierarchical scheme to evaluate the integral over the source redshift separately, breaking down the high-dimensional integral into an integral with a lower dimension and smaller degeneracy among variables that can be computed using Monte Carlo methods such as in Ref.~\cite{2004AIPC..735..395S}, plus a 1D integral over the redshift that can be efficiently evaluated. The true source parameters can be inferred after the hierarchical analysis using a Gibbs sampling algorithm.

We demonstrate the framework with examples when $\Ntrig = 1$ and $\Ntrig = 2$, respectively. We show explicitly how changing the assumed astrophysical models can alter one's interpretation on the origin of the observed signals, and sometimes can lead to smoking-gun evidence of strong lensing of GW. For the case of $\Ntrig = 2$, we also show how one can jointly constrain the total mass and the redshift of the source. Together with the improvement in localizing the source from observing two images, it helps locating the source and the lens electromagnetically \cite{Seto:2003iw, Hannuksela:2020xor}, and ultimately cross-validating the GW lensing analysis. In some cases where higher multipoles of lensed GW signals are observable, the Morse index, or the type, of the lensed images can be identified.

Currently, one of the key assumptions of the framework is that all $\Ntrig$ triggers are of astrophysical origins. This might not be the case when we perform a targeted search on subthreshold lensed counterparts \cite{Li:2019osa, McIsaac:2019use, Dai:2020tpj}, especially for a high-mass GW signal. This is because transient noise fluctuations in the detectors, or glitches, can easily mimic these short-duration high-mass signals. In this case, unfortunately, the Occam's razor is working against us as the lensed hypothesis will have a similar goodness-of-fit to the glitches compared to the not lensed hypothesis while having fewer free parameters. The assumption can be relaxed where some of the $\Ntrig$ triggers are of terrestrial origins. However, this would require a detailed study of the morphology of the glitches and we leave this to future work.

While detecting strongly lensed GW signals are exciting in its own right, it can also be used to improve our understanding on various subjects in fundamental physics, astrophysics and cosmology. For example, observing the strongly lensed GW signals from the same source multiple times effectively boosts the number of ``virtual GW detectors'', and hence can be used to better constrain the polarization contents of the GW signals and test the general relativity \cite{Goyal:2020bkm}. Strongly lensed GW signals also allow us to probe further to study for example stellar environments at higher redshift through the spin alignment of the lensed binaries \cite{Stevenson:2017dlk}, and origins of the binaries at high redshift \cite{Nakamura:2016hna, Koushiappas:2017kqm}. With the statistical framework presented in this \this{}, which directly ingests models from the astrophysics community, the GW data analysis community can work closer together with the astrophysical modeling community on finding lensed GWs and extracting more science from them. 
\acknowledgments
The authors would like to thank Tjonnie G.~F.~Li, Will Farr, Masamune Oguri, Alan Weinstein, Yanbei Chen, Katerina Chatziioannou, Colm Talbot, and Ken K.~Y.~Ng for the discussion and the help when preparing this \this.
R.~K.~L.~L.~ acknowledges support from the Croucher Foundation. R.~K.~L.~L~ also acknowledges support from NSF Awards No.~PHY-1912594, No.~PHY-2207758. I.~M.~H.~ is supported by the NSF Graduate Research Fellowship Program under Grant No.~DGE-17247915.
The computations presented here were conducted on the Caltech High Performance Cluster partially supported by a grant from the Gordon and Betty Moore Foundation. I.~M.~H.~ also acknowledges support from NSF Awards No.~PHY-1607585, No.~PHY-1912649, No.~PHY-1806990, and No.~PHY-2207728. The authors are also grateful for computational resources provided by the LIGO Laboratory and supported by National Science Foundation Grants PHY-0757058 and PHY-0823459. Figs.~ \ref{fig:not_lensed_hyp_data_gen}, \ref{fig:lensed_hyp_data_gen}, \ref{fig:not_lensed_hyp_data_gen_BBH}, and \ref{fig:lensed_hyp_data_gen_BBH} were generated using \texttt{BayesNet}.
Figures \ref{fig:Corner_highmass_logunif_mmax300}, \ref{fig:Corner_detframe60}, and \ref{fig:Corner_single_highmass_BBH} were generated using \texttt{corner.py} \cite{corner}.
This is LIGO Document No.~\LIGODCCNumber.
 
\appendix
\section{Full derivation of the probability densities of observing a set of data under various hypotheses}
\label{app:Full_derivation}
In this section, we give the full derivation for the probability densities of observing a set of data $\bm{\data}$ under the lensed hypothesis $\lensedhyp$ and the not-lensed hypothesis $\notlensedhyp$, accounting for both the astrophysical information on sources and lenses, and well as selection effects. This would in turn allow us to write down the expression for the (proper) Bayes factor and hence the posterior odds for identifying strongly lensed gravitational-wave signals. The derivations below follow Ref.~\cite{2019MNRAS.486.1086M} closely.

\subsection{Under the not-lensed hypothesis $\notlensedhyp$}
Now let us find the expression for the probability density $p(\bm{\data}|\notlensedhyp, \bm{\popparam}, \mathcal{R})$ under the not-lensed hypothesis $\notlensedhyp$.
Since each of the observed data in the data set $\bm{\data} = \{ \data^{(i)}\}_{i=1}^{i=\Ntrig}$ is simply one random draw from the population distribution $\ppop$, we can factorize the probability density into $\Ntrig$ terms as
\begin{equation}
\begin{aligned}
\label{eq:factorized_likelihood_not_lensed}
	p(\bm{\data}|\notlensedhyp, \bm{\popparam}, \mathcal{R}) \propto \prod_{i=1}^{\Ntrig} p(\data^{(i)}|\notlensedhyp, \bm{\bm{\popparam}}, \mathcal{R}).
\end{aligned}
\end{equation}
Using the marginalization rule, we can write the term $p(\data^{(i)}|\notlensedhyp, \bm{\popparam}, \mathcal{R})$ as
\begin{equation}
\label{eq:event-level_evidence_not_lensed}
\begin{aligned}
	& p(\data^{(i)}|\notlensedhyp, \bm{\popparam}, \mathcal{R}) \\
	& \propto \int \diff \bm{\eventparam}^{(i)} \; p(\data^{(i)}|\bm{\eventparam}^{(i)}) \; \ppop(\bm{\eventparam}^{(i)}|\notlensedhyp, \bm{\popparam}, \mathcal{R}),
\end{aligned}
\end{equation}
where this integral can be estimated using nested sampling \cite{2004AIPC..735..395S} in a so-called \gls{PE} run.
In a \gls{PE} run, the goal is to obtain a set of posterior samples for $\bm{\eventparam}^{(i)}$ that follow the posterior distribution
\begin{equation}
\label{eq:standard_pe_posterior}
	p(\bm{\eventparam}^{(i)}|\data^{(i)}) = \frac{p(\data^{(i)}|\bm{\eventparam}^{(i)})p_{\rm PE}(\bm{\eventparam}^{(i)})}{Z^{(i)}},
\end{equation}
where $p_{\rm PE}$ is the sampling prior distribution used in that particular \gls{PE} run, and $Z^{(i)}$ is the evidence under the particular signal hypothesis in that PE (or equivalently a normalization constant). Note that we can rearrange Eq.~\eqref{eq:event-level_evidence_not_lensed} to get
\begin{equation}
\begin{aligned}
\label{eq:reweighing}
	& p(\data^{(i)}|\notlensedhyp, \bm{\popparam}, \mathcal{R}) \\
	& \propto \int \diff \bm{\eventparam}^{(i)} \; p(\data^{(i)}|\bm{\eventparam}^{(i)}) \; \ppop(\bm{\eventparam}^{(i)}|\notlensedhyp, \bm{\popparam}, \mathcal{R}) \\
	& = \int \diff \bm{\eventparam}^{(i)} \; p(\bm{\eventparam}^{(i)}|\data^{(i)}) \; Z^{(i)} \frac{\ppop(\bm{\eventparam}^{(i)}|\notlensedhyp, \bm{\popparam}, \mathcal{R})}{p_{\rm PE}(\bm{\eventparam}^{(i)})} \\
	& \approx \langle Z^{(i)} \frac{\ppop(\bm{\eventparam}^{(i)}|\notlensedhyp, \bm{\popparam}, \mathcal{R})}{p_{\rm PE}(\bm{\eventparam}^{(i)})} \rangle,
\end{aligned}
\end{equation}
where $\langle \cdots \rangle$ denotes an average over posterior samples for $\bm{\eventparam}^{(i)}$.
This means that we can obtain the unnormalized probability density as
\begin{equation}
	p(\bm{\data}|\notlensedhyp, \bm{\popparam}, \mathcal{R}) \propto \prod_{i=1}^{\Ntrig}  Z^{(i)} \langle \frac{\ppop(\bm{\eventparam}^{(i)}|\notlensedhyp, \bm{\popparam}, \mathcal{R})}{p_{\rm PE}(\bm{\eventparam}^{(i)})} \rangle
\end{equation}

Now, we will need to obtain the normalization constant, denoted as $\tilde{\alpha}$, by requiring that when summed over all observable data sets Eq.~\eqref{eq:factorized_likelihood_not_lensed} sums to unity, i.e.,
\begin{equation}
	\frac{1}{\tilde{\alpha}} \int_{\text{all obs. data}} \diff^{\Ntrig} \bm{\data} \; \prod_{i=1}^{\Ntrig} p(\data^{(i)}|\notlensedhyp, \bm{\popparam}, \mathcal{R}) = 1.
\end{equation}
Since the likelihood function $p(\data^{(i)}|\notlensedhyp, \bm{\popparam}, \mathcal{R})$ is independent of each other, we have
\begin{equation}
\begin{aligned}
	\tilde{\alpha} & = \int_{\text{all obs. data}} \diff^{\Ntrig} \bm{\data} \; \prod_{i=1}^{\Ntrig} p(\data^{(i)}|\notlensedhyp, \bm{\popparam}, \mathcal{R})  \\
	& = \prod_{i=1}^{\Ntrig} \underbrace{\int_{\text{all obs. data}} \diff \data^{(i)} \; p(\data^{(i)}| \bm{\popparam}, \mathcal{R})}_{\text{Selection function } \alpha(\bm{\popparam}, \mathcal{R})} \\
	& = \alpha(\bm{\popparam}, \mathcal{R})^{\Ntrig},
\end{aligned}
\end{equation}
where $\alpha(\bm{\popparam}, \mathcal{R}) \equiv \int_{\text{all obs. data}} \diff \data^{(i)} \;  p(\data^{(i)}|\bm{\popparam}, \mathcal{R})$ is known as the selection function or detectable fraction for the population model parametrized by $\bm{\popparam}$ with a merger rate density $\mathcal{R}$. We can write the selection function $\alpha(\bm{\popparam}, \mathcal{R})$ as an integral over all observable data and the event-level parameters $\bm{\theta}^{(i)}$ as
\begin{equation}
\begin{aligned}
	\alpha(\bm{\popparam}, \mathcal{R}) & = \int_{\text{all obs. data}} \diff \data^{(i)} \;  \\
	& \times \int \diff \bm{\eventparam}^{(i)} \; p(\data^{(i)}|\bm{\eventparam}^{(i)}) \; \ppop(\bm{\eventparam}^{(i)}| \notlensedhyp, \bm{\popparam}, \mathcal{R}) \\
	& = \int \diff \bm{\eventparam}^{(i)} \; \underbrace{\left( \int_{\text{all obs. data}} \diff \data^{(i)} \; p(\data^{(i)}|\bm{\eventparam}^{(i)})\right)}_{\text{detection probability } p_{\rm det}(\bm{\eventparam} = \bm{\eventparam}^{(i)})} \; \\
	& \times \ppop(\bm{\eventparam}^{(i)}|\notlensedhyp, \bm{\popparam}, \mathcal{R}) \\
	& = \int \diff \bm{\eventparam}^{(i)} \;	p_{\rm det}(\bm{\eventparam}^{(i)}) \; \ppop(\bm{\eventparam}^{(i)}| \notlensedhyp, \bm{\popparam}, \mathcal{R}),
\end{aligned}
\end{equation}
where $p_{\rm det}(\bm{\eventparam})$ is known as the detection probability. The detection probability can be obtained semianalytically, or empirically estimated by performing an injection campaign \cite{Tiwari:2017ndi, 2019RNAAS...3...66F, LIGOScientific:2018jsj, Abbott:2020gyp}. In Appendix~\ref{app:Evaluation_of_alpha}, we give more details on the numerical computation of the selection function $\alpha$.

To conclude, the full expression of the probability of observing the data set $\bm{\data}$ under the not-lensed hypothesis is given by
\begin{equation}
\begin{aligned}
\label{eq:not_lensed_hyp_data_prob}
	& p(\bm{\data}|\notlensedhyp, \bm{\popparam}, \mathcal{R}) \\
	& = \frac{1}{\alpha(\bm{\popparam}, \mathcal{R})^{\Ntrig}} \prod_{i=1}^{\Ntrig} p(\data^{(i)}|\notlensedhyp, \bm{\popparam}, \mathcal{R}).
\end{aligned}
\end{equation}

\subsection{Under the lensed hypothesis $\lensedhyp$}
For the lensed hypothesis $\lensedhyp$, we only make \emph{one} random draw from the source population distribution $\srcpop$, and $\Ntrig$ random draws of the parameters of the lensed images from the lens population distribution $\lenspop$.

Using the marginalization rule, we write the probability of observing the data set $\bm{\data}$ under this hypothesis as
\begin{equation}
\begin{aligned}
& p(\bm{\data}|\lensedhyp, \bm{\popparam}, \mathcal{R}, \bm{\lensparam}) \\
& \propto \int \diff\bm{\eventparam}^{(1)}\cdots \diff \bm{\eventparam}^{(\Ntrig)} \; p(\bm{\eventparam}^{(1)}, ..., \bm{\eventparam}^{(\Ntrig)}| \lensedhyp, \bm{\popparam}, \mathcal{R}, \bm{\lensparam}) \\
& \times p(\bm{\data}|\bm{\eventparam}^{(1)}, ..., \bm{\eventparam}^{(\Ntrig)}, \lensedhyp, \bm{\popparam}, \mathcal{R}, \bm{\lensparam}).
\end{aligned}
\end{equation}
Recall that we partition the event-level parameters $\bm{\eventparam}^{(i)}$ into common parameters $\comparam{}$ and independent parameters $\idparam{}$. As a result, we can write the probability density as
\begin{equation}
\begin{aligned}
& p(\bm{\data}|\lensedhyp, \bm{\popparam}, \mathcal{R}, \bm{\lensparam}) \\ 
& \propto \int \diff\comparam{(1)} \; \diff\idparam{(1)} \cdots \diff\comparam{(\Ntrig)} \diff\idparam{(\Ntrig)} \; p(\bm{\data}|\bm{\eventparam}^{(1)}, ..., \bm{\eventparam}^{(\Ntrig)}) \\
& \times p(\comparam{(1)},...,\comparam{(\Ntrig)}| \lensedhyp, \bm{\popparam}, \mathcal{R}, \bm{\lensparam}) \\
& \times p(\idparam{(1)},...,\idparam{(\Ntrig)}|\comparam{(1)},...,\comparam{(\Ntrig)}, \lensedhyp, \bm{\popparam}, \mathcal{R}, \bm{\lensparam}).
\end{aligned}
\end{equation}
We expect the common parameters to be the same across the $\Ntrig$ events, and that the independent parameters $\idparam{}$ to be distributed according to the same lens population distribution, i.e., $\idparam{} \sim \pid(\idparam{}|\bm{\lensparam})$. We also assume they are independent of the common parameters. Therefore, we have
\begin{equation}
\begin{aligned}
& p(\bm{\data}|\lensedhyp, \bm{\popparam}, \mathcal{R}, \bm{\lensparam}) \\
& \propto \int \diff\comparam{(1)} \; \diff\idparam{(1)} \cdots \diff\comparam{(\Ntrig)} \; \diff\idparam{(\Ntrig)} \; p(\bm{\data}|\bm{\eventparam}^{(1)}, ..., \bm{\eventparam}^{(\Ntrig)}) \\
& \times \pid(\idparam{(1)},...,\idparam{(\Ntrig)}|\bm{\lensparam}) \pcom(\comparam{(1)}|\bm{\popparam}, \mathcal{R}) \\
& \times \delta(\comparam{(2)} - \comparam{(1)}) \cdots \delta(\comparam{(\Ntrig)} - \comparam{(1)}),
\end{aligned}
\end{equation}
where we impose the $\left( \Ntrig - 1 \right)$ Dirac-delta distribution to enforce the common parameters to be the same across the $\Ntrig$ events, reducing the dimension of the integral. Now the expression simplifies to
\begin{equation}
\begin{aligned}
& p(\bm{\data}|\lensedhyp, \bm{\popparam}, \mathcal{R}, \bm{\lensparam}) \\
& \propto \int \diff\comparam{(1)} \; \diff\idparam{(1)} \cdots \diff\idparam{(\Ntrig)} p(\bm{\data}|\idparam{(1)}, ..., \idparam{(\Ntrig)}, \comparam{(1)}) \\
& \times \pid(\idparam{(1)},...,\idparam{(\Ntrig)}|\bm{\lensparam}) \pcom(\comparam{(1)}|\bm{\popparam}, \mathcal{R}).
\end{aligned}
\end{equation}
Note that the joint-likelihood function can still be factorized as
\begin{equation}
\begin{aligned}
& p(\boldvec{\data}|\idparam{(1)}, ..., \idparam{(\Ntrig)}, \comparam{(1)}) & \\
& \propto p(\data^{(1)}|\idparam{(1)},\comparam{(1)}) \cdots p(\data^{(\Ntrig)}|\idparam{(\Ntrig)},\comparam{(1)}).
\end{aligned}
\end{equation}
Therefore, we have
\begin{equation}
\begin{aligned}
& p(\bm{\data}|\lensedhyp, \bm{\popparam}, \mathcal{R}, \bm{\lensparam}) \\
& \propto \int \diff\comparam{(1)} \; \diff\idparam{(1)} \cdots \diff\idparam{(\Ntrig)} \; p(\data^{(1)}|\idparam{(1)},\comparam{(1)}) \cdots \\
& \times p(\data^{(\Ntrig)}|\idparam{(\Ntrig)},\comparam{(1)}) \\
& \times\pid(\idparam{(1)},...,\idparam{(\Ntrig)}|\bm{\lensparam}) \pcom(\comparam{(1)}|\bm{\popparam}, \mathcal{R}).
\end{aligned}
\end{equation}

To evaluate the evidence integral above, one can use nested sampling \cite{2004AIPC..735..395S} to perform a joint-parameter estimation across the $\Ntrig$ events with the chosen source and lens population distribution, as well as Dirac-delta prior distribution to enforce the common parameters to be the same across the events. In particular, we choose the likelihood function $\mathcal{L}_{\text{joint-PE}}$ in a joint-PE run as
\begin{equation}
\begin{aligned}
	\mathcal{L}_{\text{joint-PE}} & = p(\data^{(1)}|\idparam{(1)},\comparam{(1)}) \cdots p(\data^{(\Ntrig)}|\idparam{(\Ntrig)},\comparam{(1)}) \\
	& = \prod_{j=1}^{\Ntrig} p(\data^{(j)}|\idparam{(j)},\comparam{(1)}),
\end{aligned}
\end{equation}
and with the joint prior distribution $p_{\text{joint-PE}} = p_{\text{joint-PE}}(\idparam{(1)},...,\idparam{(\Ntrig)}, \comparam{(1)})$
with parameters $\{ \comparam{(1)}, \idparam{(1)},...,\idparam{(\Ntrig)}\}$ being sampled over in the joint-PE run.

Still, we will need to find the overall normalization constant $\beta$ by again requiring when summed over all observable data sets that
\begin{equation}
\begin{aligned}
	1 = & \int_{\text{all obs. data}} \diff^{N} \bm{\data} \; p(\bm{\data}|\lensedhyp, \bm{\popparam}, \mathcal{R}, \bm{\lensparam}) \\
	= & \frac{1}{\beta} \int_{\text{all obs. data}} \diff^{N} \bm{\data} \int \diff\comparam{(1)} \; \diff\idparam{(1)} \cdots \diff\idparam{(\Ntrig)} \; \\
	& \Bigl[\; p(\data^{(1)}|\idparam{(1)},\comparam{(1)}) \cdots p(\data^{(\Ntrig)}|\idparam{(\Ntrig)},\comparam{(1)}) \\
	& \times \pid(\idparam{(1)},...,\idparam{(\Ntrig)}|\bm{\lensparam}) \pcom(\comparam{(1)}|\bm{\popparam}, \mathcal{R}) \; \Bigr] \\
	\beta = & \int_{\text{all obs. data}} \diff^{N} \bm{\data} \int \diff\comparam{(1)} \; \diff\idparam{(1)} \cdots \diff\idparam{(\Ntrig)} \; \\
	& p(\data^{(1)}|\idparam{(1)},\comparam{(1)}) \cdots p(\data^{(\Ntrig)}|\idparam{(\Ntrig)},\comparam{(1)}) \\
	& \times \pid(\idparam{(1)},...,\idparam{(\Ntrig)}|\bm{\lensparam}) \pcom(\comparam{(1)}|\bm{\popparam}, \mathcal{R}).
\end{aligned}
\end{equation}
The normalization constant $\beta = \beta(\bm{\popparam}, \mathcal{R}, \bm{\lensparam})$ can be viewed as a function of the population models. We can further write this ``lensing selection function'' as
\begin{equation}
\begin{aligned}
	 & \beta(\bm{\popparam}, \mathcal{R}, \bm{\lensparam}) \\
	 & = \int_{\text{all obs. data}} \diff^{N} \boldvec{\data} \int \diff\comparam{(1)} \; \diff\idparam{(1)} \cdots \diff\idparam{(\Ntrig)} \; \\ 
	& \times p(\data^{(1)}|\idparam{(1)},\comparam{(1)}) \cdots p(\data^{(\Ntrig)}|\idparam{(\Ntrig)},\comparam{(1)})  \\
	& \times \pid(\idparam{(1)},...,\idparam{(\Ntrig)}|\bm{\lensparam}) \pcom(\comparam{(1)}|\bm{\popparam}, \mathcal{R}) \\
	& = \int \diff\comparam{(1)} \; \diff\idparam{(1)} \cdots \diff\idparam{(\Ntrig)} \; \\
	& \times \underbrace{\left( \int_{\text{all obs. data}} d \data^{(1)} \; p(\data^{(1)}|\idparam{(1)},\comparam{(1)}) \right)}_{p_{\rm det}(\bm{\eventparam} = \comparam{(1)} \cup \idparam{(1)})} \\
	& \times \cdots \\
	& \times \underbrace{\left( \int_{\text{all obs. data}} \diff \data^{(\Ntrig)} \; p(\data^{(\Ntrig)}|\idparam{(\Ntrig)},\comparam{(1)}) \right)}_{p_{\rm det}(\bm{\eventparam} = \comparam{(1)} \cup \idparam{(\Ntrig)})}. \\
	& \times \pid(\idparam{(1)},...,\idparam{(\Ntrig)}|\bm{\lensparam}) \pcom(\comparam{(1)}|\bm{\popparam}, \mathcal{R}). \\
& = \int \diff\comparam{(1)} \; \diff\idparam{(1)} \cdots \diff\idparam{(\Ntrig)} \; \left[ \prod_{j=1}^{N} p_{\rm det}(\bm{\eventparam} = \comparam{(j)} \cup \idparam{(1)}) \right] \\
& \times \pid(\idparam{(1)},...,\idparam{(\Ntrig)}|\bm{\lensparam}) \pcom(\comparam{(1)}|\bm{\popparam}, \mathcal{R}).
\end{aligned}
\end{equation}
Therefore, in order to evaluate the lensing selection function $\beta$, we will need to perform a (possibly Monte Carlo) integration, evaluated at the chosen population models.  In Appendix~\ref{app:Evaluation_of_beta}, we give more details on the numerical computation of the selection function $\beta$.

In summary, the full expression of the probability of observing the data set $\bm{\data}$ under the lensed hypothesis is
\begin{equation}
\begin{aligned}
\label{eq:lensed_hyp_data_prob}
	& p(\bm{\data}|\lensedhyp, \bm{\popparam}, \mathcal{R}, \bm{\lensparam}) \\ & =  \frac{1}{\beta(\bm{\popparam}, \mathcal{R}, \bm{\lensparam})} \int \diff\comparam{(1)} \; \diff\idparam{(1)} \; \cdots \diff\idparam{(\Ntrig)} \; \left[ \prod_{j=1}^{N} p(\data^{(j)}|\idparam{(j)},\comparam{(1)}) \right] \\
	& \times \pid(\idparam{(1)},...,\idparam{(\Ntrig)}|\bm{\lensparam}) \pcom(\comparam{(1)}|\bm{\popparam}, \mathcal{R}).
\end{aligned}
\end{equation}

\section{Evaluation of selection functions}
\label{app:Evaluation_of_selection_functions}
Here we explain how to numerically compute the selection functions $\alpha$ and $\beta$ under the not-lensed and the lensed hypothesis respectively. Note that the numerical values are parametrization invariant, and hence we will be using the parametrization that is more convenient in terms of computation. A common technique employed in the evaluation of high-dimensional integral, as in the case here, is Monte Carlo (MC) integration, where we randomly generate points inside the integration region for evaluation. We will be using MC integration extensively here.

\subsection{Under the not-lensed hypothesis}
\label{app:Evaluation_of_alpha}
The selection function under the not-lensed hypothesis $\alpha(\bm{\popparam}, \mathcal{R})$, is given by
\begin{equation}
\begin{aligned}
	\alpha(\bm{\popparam}, \mathcal{R}) & = \int_{\text{all obs. data}} \diff \data \;  \int \diff \bm{\eventparam} \; p(\data|\bm{\eventparam}) \; \ppop(\bm{\eventparam}|\bm{\popparam}, \mathcal{R}) \\
	& = \int \diff \bm{\eventparam} \; \underbrace{\left[ \int \diff \data \;  \Theta(\rho - \rho_{\rm th}) p(\data|\bm{\eventparam}) \right]}_{p_{\rm det}(\bm{\eventparam})} \ppop(\bm{\eventparam}|\bm{\popparam}, \mathcal{R}) \\
	& = \int \diff \bm{\eventparam} \; p_{\rm det}(\bm{\eventparam}) \ppop(\bm{\eventparam}|\bm{\popparam}, \mathcal{R}), \\
\end{aligned}
\end{equation}
where we have chosen the network SNR threshold $\rho_{\rm th}$ to be $12$ \cite{Gerosa:2020pgy}, and $\Theta(x)$ denotes the Heaviside step function.

We can evaluate this integral using Monte Carlo integration. If we have $N_{\rm MC}$ samples of $\bm{\eventparam}$ drawn from the distribution $\ppop(\bm{\eventparam}|\bm{\popparam}, \mathcal{R})$, then we can simply approximate the selection function as
\begin{equation}
		\alpha(\bm{\popparam}, \mathcal{R}) \approx \frac{1}{N_{\rm MC}} \sum_{i=1}^{N_{\rm MC}} p_{\rm det}(\bm{\eventparam}_{i}).
\end{equation}

However, for most of the time, it is not trivial to generate samples from $\ppop(\bm{\eventparam}|\bm{\popparam}, \mathcal{R})$. We can generate a set of fiducial samples $\left\{ \bm{\eventparam} \right\}_{i}$ that instead follow another distribution $q(\bm{\eventparam})$ that we can sample easily. This is known as importance sampling (where $q$ should be chosen such that it is nonvanishing wherever $\ppop$ is also nonvanishing). The selection function $\alpha$ can be calculated as
\begin{equation}
	\alpha(\bm{\popparam}, \mathcal{R}) \approx \frac{1}{N_{\rm MC}} \sum_{i=1}^{N_{\rm MC}} \left[ p_{\rm det}(\bm{\eventparam}_{i}) \frac{\ppop(\bm{\eventparam}_{i}|\bm{\popparam}, \mathcal{R})}{q(\bm{\eventparam}_{i})} \right]
\end{equation}

\subsection{Under the lensed hypothesis}
\label{app:Evaluation_of_beta}
The selection function under the lensed hypothesis $\beta(\bm{\popparam}, \mathcal{R}, \bm{\lensparam})$ is given by
\begin{equation}
\begin{aligned}
& \beta(\bm{\popparam}, \mathcal{R}, \bm{\lensparam}) \\
& = \int \diff\comparam{(1)} \; \diff\idparam{(1)} \cdots \diff\idparam{(\Ntrig)} \; \left[ \prod_{j=1}^{N} p_{\rm det}(\bm{\eventparam} = \comparam{(j)} \cup \idparam{(1)}) \right] \\
& \times \pid(\idparam{(1)},...,\idparam{(\Ntrig)}|\bm{\lensparam}) \pcom(\comparam{(1)}|\bm{\popparam}, \mathcal{R}).
\end{aligned}
\end{equation}
Here we ignore any kind of observational selection effects, for example due to detector down-time \cite{Chen:2016luc} or finite observation period \cite{Li:2018prc}.
Suppose we are using the parametrization $\comparam{} = \left\{ m_{1}, m_{2}, \bm{\chi}_{1}, \bm{\chi}_{2}, \alpha, \delta, \psi, \phi_{\rm c} \right\}$ and $\idparam{} = \left\{ \mu^{(1)}, \mu^{(2)}, \Xi^{(1)}, \Xi^{(2)}, z \right\}$. We also ignore the phasing effect due to lensing to the detectability of images \cite{Wang:2021kzt}. In particular, we can separate the integration over the source redshift $z$
\begin{equation}
\begin{aligned}
	& \beta(\bm{\popparam}, \mathcal{R}, \bm{\lensparam}) \\
	& =  \int \diff z \; \lenspop(z|\lensedhyp, \mathcal{R}) \left\{ \int \diff \left\{ m_{1}, m_{2}, \bm{\chi}_{1}, \bm{\chi}_{2}, \alpha, \delta, \psi, \phi_{\rm c} \right\} \right. \\
	& \left. \times \; \srcpop(m_{1}, m_{2}, \bm{\chi}_{1}, \bm{\chi}_{2}|\bm{\popparam}) p_{\rm ext}(\alpha, \delta, \psi, \phi_{\rm c}) \right. \\
	& \left. \times \int \diff \left\{ \mu^{(1)}, \mu^{(2)} \right\} \; \lenspop(\mu^{(1)}, \mu^{(2)}| \bm{\lensparam}) \; \left[ p_{\rm det}(\bm{\eventparam}^{(1)}) p_{\rm det}(\bm{\eventparam}^{(2)}) \right] \right\},
\end{aligned}
\end{equation}
where the dependence on the image types $\Xi^{(1)}, \Xi^{(2)}$ are trivially integrated over since $\int \diff \Xi^{(i)} \lenspop(\Xi^{(i)}) = 1$.

\begin{figure}[h!]
\centering
	\includegraphics[width=\columnwidth]{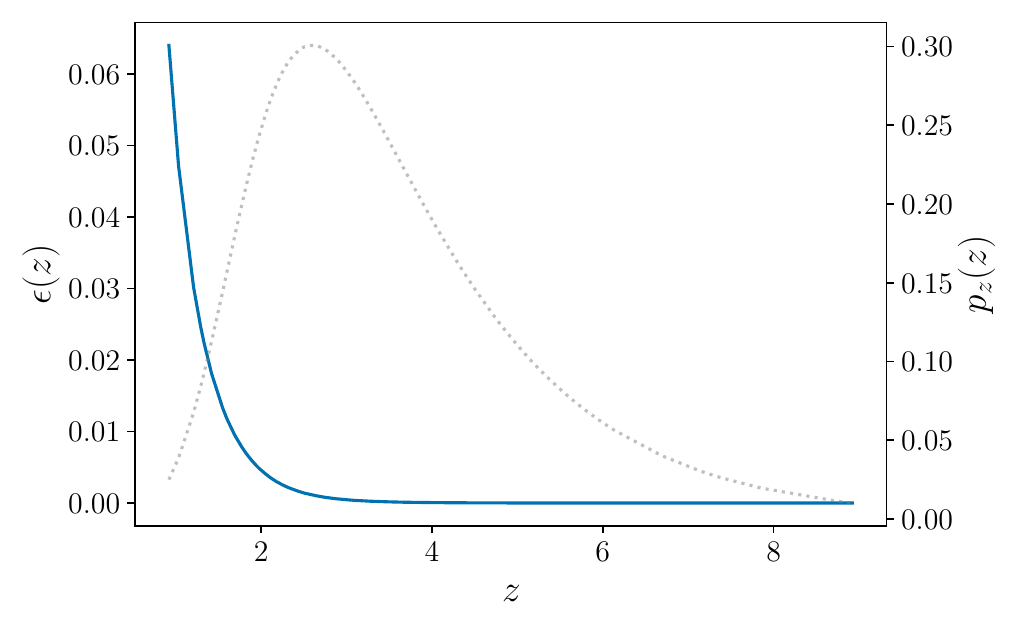}
	\caption{\label{fig:epsilon_z}The inner integral $\epsilon(z)$ (solid line) as a function of the source redshift $z$ evaluated using Monte Carlo integration. We see that $\epsilon(z)$ decreases as we go further in $z$. However, the contribution of $\epsilon(z)$ to the $\beta$ selection function integral is weighted by $p_z(z)$ (dotted line) which peaks at $z \sim 3$ for the particular merger rate density model and lens model we used in the calculation (cf. Sec.~\ref{sec:BBH_pair_lensing}).}
\end{figure}

Note that the event-level parameters $\bm{\eventparam}^{(i)}$ consist of $\bm{\eventparam}^{(i)} = \left\{ m_{1}, m_{2}, \bm{\chi}_{1}, \bm{\chi}_{2}, \alpha, \delta, \psi, \phi_{\rm c}, \mu^{(i)}, z \right\}$. This means that we can evaluate the inner integral enclosed by the curly brackets first using Monte Carlo integration. Again, if we have $N_{\rm MC}$ samples of $\bm{\eventparam}$ drawn from the joint distribution $\srcpop(m_{1}, m_{2}, \bm{\chi}_{1}, \bm{\chi}_{2}| \bm{\popparam})p_{\rm ext}(\alpha, \delta, \psi, \phi_{\rm c})$ and $\lenspop(\mu^{(i)}|\bm{\lensparam})$ with a given redshift $z$, then we can simply approximate the inner integral $\epsilon(z)$ as
\begin{equation}
	\epsilon(z) \approx \frac{1}{N_{\rm MC}} \sum_{i=1}^{N_{\rm MC}} \left[ \prod_{j=1}^{N} p_{\rm det}(\bm{\eventparam}^{(j)}_{i}|z) \right].
\end{equation}
Figure~\ref{fig:epsilon_z} shows how $\epsilon(z)$ depends on the redshift $z$. We see that $\epsilon(z)$ decreases as we go higher in $z$. However, the contribution of $\epsilon(z)$ to the $\beta$ selection function integral is weighted by $p_z(z)$ (dotted line) which peaks at $z \sim 3$ for the particular merger-rate density model and lens model we used in the calculation (cf. Sec.~\ref{sec:BBH_pair_lensing}).

We can employ the same Monte Carlo integration technique to evaluate the outer integral over $z$. Suppose we have $N_{z}$ samples of $z \sim p_{z}(z)$, then we can approximate the entire $\beta(\bm{\popparam}, \mathcal{R}, \bm{\lensparam})$ integral as
\begin{equation}
	\beta(\bm{\popparam}, \mathcal{R}, \bm{\lensparam}) \approx \frac{1}{N_{z}} \sum_{k=1}^{N_{z}} \frac{1}{N_{\rm MC}} \sum_{i=1}^{N_{\rm MC}} \left[ \prod_{j=1}^{N} p_{\rm det}(\bm{\eventparam}^{(j)}_{i}|z_{k}) \right].
\end{equation}

\section{Derivation of the arrival time probability density function under the not-lensed hypothesis}
\label{app:Time_delay_not_lensed}
Here we give a derivation of the arrival time probability density function under the not-lensed hypothesis we used in the main text, namely
\begin{equation}
\label{eq:arrival_time_prior_poisson}
	p(\Delta t|\notlensedhyp) = \frac{2}{T_{\rm obs}} \left( 1 - \frac{\Delta t}{T_{\rm obs}} \right).
\end{equation}
Given that we have observed $N_{\rm obs}$ events within the interval $(0, T_{\rm obs}]$, if we assume that the arrival of the events follows a Poisson process with a mean rate $r$, then we can write down the joint probability distribution of the $N_{\rm obs}$ ordered arrival times $t_{1}, t_{2}, \dots, t_{N_{\rm obs}}$ (such that $t_1 < t_2 < \dots $) conditioned that we have observed $N_{\rm obs}$ events as
\begin{widetext}
\begin{equation}
\begin{aligned}
	& p(t_{1}, t_{2}, \dots, t_{N_{\rm obs}}, t_1 < t_2 < \dots < t_{N_{\rm obs}}|N_{\rm obs}, r) \\
	& = \frac{p(t_{1}, t_{2}, \dots, t_{N_{\rm obs}}, t_1 < t_2 < \dots < t_{N_{\rm obs}}, N_{\rm obs}| r)}{p(N_{\rm obs}| r)} \\
	& = \frac{r\exp(-rt_1) \; r\exp\left[-r(t_2 - t_1)\right] \; \dots \; r\exp\left[-r(t_{N_{\rm obs}} - t_{N_{\rm obs}-1})\right] \; \exp\left[ -r(T_{\rm obs} - t_{N_{\rm obs}})\right]}{(rT_{\rm obs})^{N_{\rm obs}} \exp(-rT_{\rm obs})/N_{\rm obs}!} \\
	& = \frac{N_{\rm obs}!}{T_{\rm obs}^{N_{\rm obs}}},
\end{aligned}
\end{equation}
\end{widetext}
where the second line uses the fact that for a Poisson process one can partition the time interval into many smaller chunks where each chunk still follows the same Poisson process, and that the interarrival time follows an exponential distribution. The factor $\exp\left[ -r(T_{\rm obs} - t_{N_{\rm obs}})\right]$ is due to the requirement that there is no event between $( t_{N_{\rm obs}}, T_{\rm obs} ]$. Note that there are exactly $N_{\rm obs}!$ combinations of unordered arrival times $\left\{ t_{1}, t_{2}, \dots, t_{N_{\rm obs}} \right\}$ that would lead to the same ordered times. Therefore the joint probability distribution of the $N_{\rm obs}$ unordered arrival times $\left\{ t_{1}, t_{2}, \dots, t_{N_{\rm obs}} \right\}$ is simply
\begin{equation}
	p(t_{1}, t_{2}, \dots, t_{N_{\rm obs}}|N_{\rm obs}, r) = \left( \frac{1}{T_{\rm obs}} \right)^{N_{\rm obs}}.
\end{equation}

Note that we are interested in the probability distribution of the time delay between arbitrary two events among these $N_{\rm obs} \geq 2$ events. If we consider the unordered set of arrival times, without loss of generality we can assume that the arrival times corresponding to the two events are $t_{1}$ and $t_{2}$ respectively. The marginalization over $\left\{ t_{3}, t_{4}, \dots, t_{N_{\rm obs}} \right\}$ is trivial since the joint probability density does not depend on $t_{i}$, and the marginalized density is just
\begin{equation}
	p(t_{1}, t_{2}|N_{\rm obs}, r) = \left( \frac{1}{T_{\rm obs}} \right)^{2}.
\end{equation}
Note that since $0 < t_{1}, t_{2} \leq T_{\rm obs}$, this probability distribution is simply two uniform distributions multiplied together, i.e.
\begin{equation}
	p(t_{1}|N_{\rm obs}, r) = p(t_{2}|N_{\rm obs}, r) = \left( \frac{1}{T_{\rm obs}} \right).
\end{equation}

\begin{figure}[h]
\centering
\includegraphics[width=\columnwidth]{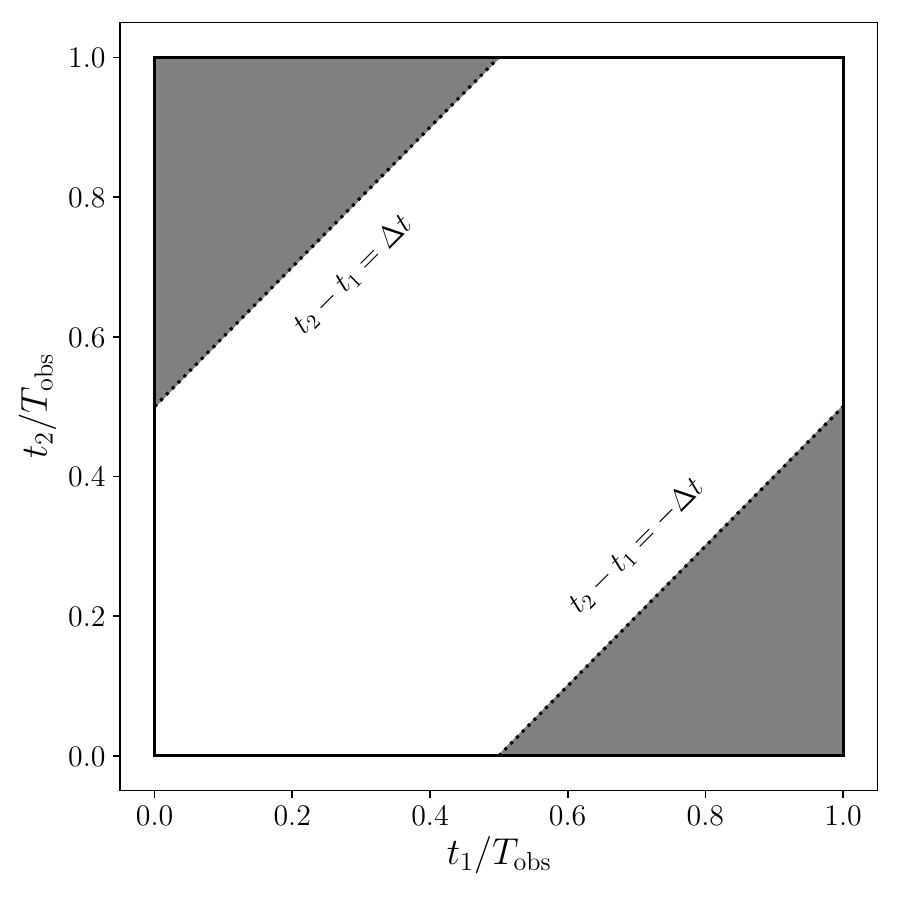}
\caption{\label{fig:integration_region}A visualization of the integration region over the $t_{1}/T_{\rm obs} - t_{2}/T_{\rm obs}$ plane to obtain the distribution $p(\Delta t|\notlensedhyp)$. The condition that $\Delta t > 0$ carves out two triangles of equal area $(1 - \Delta t/T_{\rm obs})^2/2$ from the unit square. In the figure, we put $\Delta t/T_{\rm obs} = 1/2$ for demonstration.}
\end{figure}

Let us define a new variable $\Delta t = |t_{2} - t_{1}| > 0$. The cumulative distribution function $F(\Delta t) \equiv \Pr (|t_{2} - t_{1}| < \Delta t)$ can be found using simple coordinate geometry by noting that the support of the joint distribution of $p(t_{1}, t_{2}|N_{\rm obs}, r)$ forms a square of length $T_{\rm obs}$ (see Fig.~\ref{fig:integration_region}). If we normalize the length by $T_{\rm obs}$, the condition that $\Delta t > 0$ cuts out two triangles of area $(1 - \Delta t/T_{\rm obs})^2/2$ from the unit square. Therefore, 
\begin{equation}
\begin{aligned}
	F(\Delta t) & = \Pr ( -\Delta t < t_{2} - t_{1} < \Delta t) \\
	& = 1 - 2 \frac{(1 - \Delta t/T_{\rm obs})^2}{2}.
\end{aligned}
\end{equation}

\begin{figure}[h]
\centering
\includegraphics[width=\columnwidth]{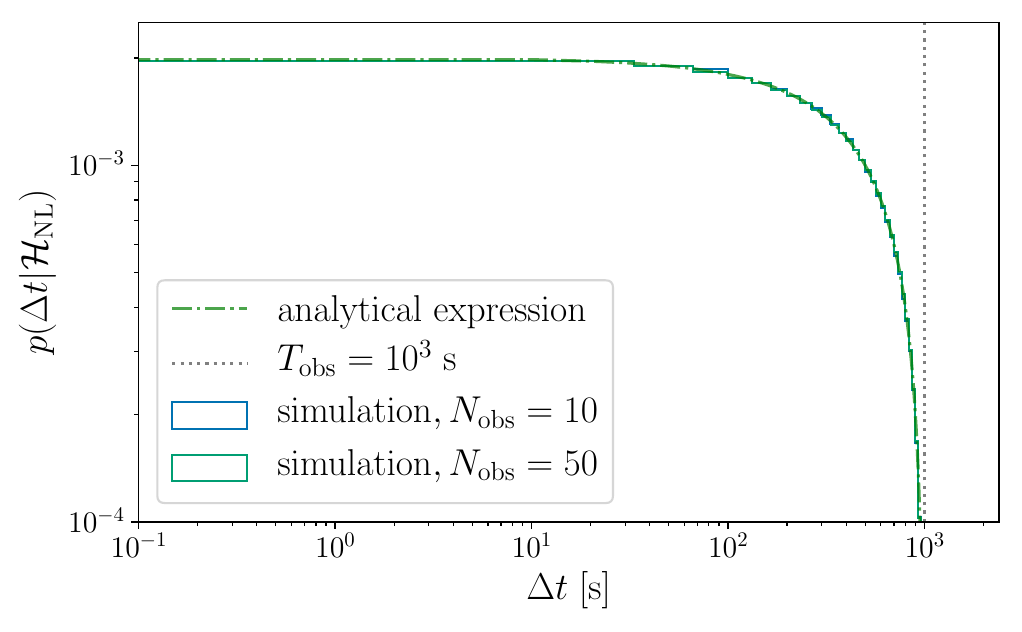}
\caption{\label{fig:arrival_time_analytical_simulation} The probability density $p(\Delta t|\notlensedhyp)$ for the time delay $\Delta t $ under the not-lensed hypothesis from an analytical expression (dot-dashed line) and from simulations (blue and green solid lines). For visualization, we set $T_{\rm obs} = 10^{3}$ s (dotted line) when evaluating the analytical expression and performing the simulations. We see that the distributions are independent of $N_{\rm obs}$ and well-described by the analytical expression in Eq.~\eqref{eq:arrival_time_prior_poisson}.}
\end{figure}

The desired probability density is a triangular distribution, i.e.,
\begin{equation}
\begin{aligned}
	p(\Delta t|\notlensedhyp) & = \frac{\diff F}{\diff \Delta t} \\
	& = \frac{2}{T_{\rm obs}} \left( 1 - \frac{\Delta t}{T_{\rm obs}} \right),
\end{aligned}
\end{equation}
where the distribution is independent of $r$ and $N_{\rm obs}$. This can be checked against simulations. Figure \ref{fig:arrival_time_analytical_simulation} shows two histograms of $\Delta t$ from simulations with $N_{\rm obs} = 10$ and $N_{\rm obs} = 50$ respectively. We can see that the distributions for these two cases are indeed the same, and are well described by Eq.~\eqref{eq:arrival_time_prior_poisson}.

\section{Derivation of the poor-man's prior on the relative magnification}
\label{app:Poorman}
Here we will derive the poor-man's prior on the relative magnification $\mu_{\rm rel} \equiv \mu^{(2)}/\mu^{(1)}$ we used in the main text, namely
\begin{equation}
\label{eq:Poor-man}
	p(\mu_{\rm rel}) =
	\begin{cases}
		\mu_{\rm rel} & \text{for } \mu_{\rm rel} \leq 1 \\
		\mu_{\rm rel}^{-3} & \text{for } \mu_{\rm rel} > 1 \\
	\end{cases},
\end{equation}
if we assume that the absolute magnification follows the probability distribution
\begin{equation}
	p(\mu) = 
	\begin{cases}
		2 \mu_{\rm min}^{2} \mu^{-3} & \text{for } \mu \geq \mu_{\rm min} \\
		0 & \text{otherwise}
	\end{cases},
\end{equation}
where $\mu_{\rm min} > 0$.

\begin{figure}[h]
\centering
\includegraphics[width=\columnwidth]{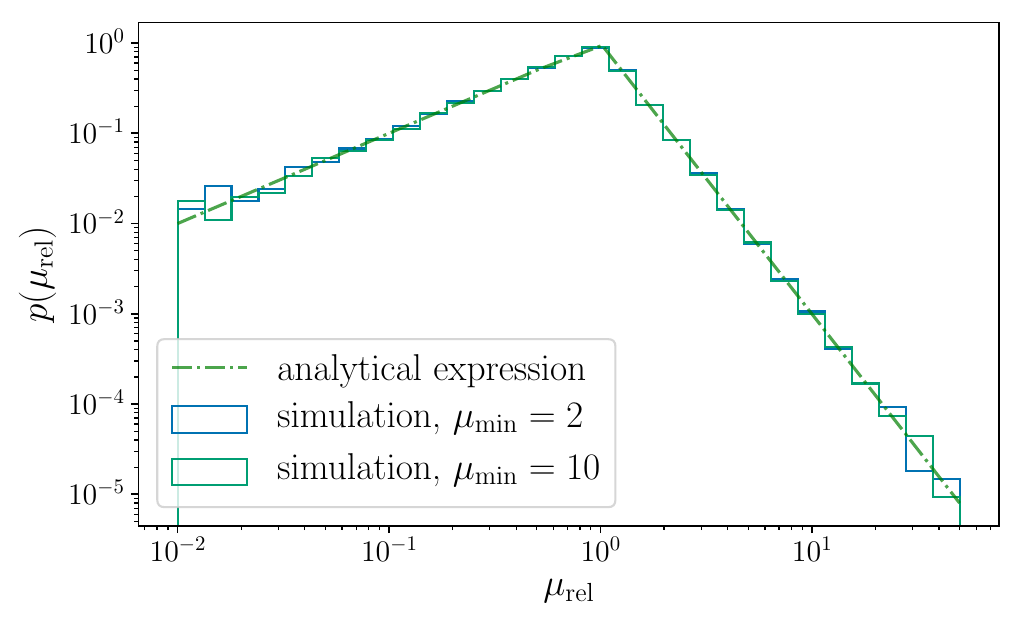}
\caption{\label{fig:Poor-man}The poor-man's prior for the relative magnification. Numerical simulations were performed with different values of $\mu_{\rm min}$ to confirm the analytical expression in Eq.~\eqref{eq:Poor-man}, and that the distribution (both the shape and normalization) does not depend on the value of $\mu_{\rm min}$.}
\end{figure}

Note that the probability distribution of the ratio of two random variables $\mu^{(1)}, \mu^{(2)}$ can be found by evaluating the integral
\begin{equation}
\begin{aligned}
	& p(\mu_{\rm rel} \equiv \mu^{(2)}/\mu^{(1)}) \\
	& = \int_{-\infty}^{+\infty} \diff \mu^{(1)} \; |\mu^{(1)}| \; p(\mu^{(1)}) \; p(\mu^{(2)} = \mu_{\rm rel}\mu^{(1)}).
\end{aligned}
\end{equation}
We can divide the integral into two cases; when $\mu_{\rm rel} > 1$ and when $\mu_{\rm rel} \leq 1$. For the former case, the condition that $\mu^{(2)} > \mu_{\rm min}$ is trivially satisfied. Therefore when $\mu_{\rm rel} > 1$, the lower limit of the integral is simply $\mu_{\rm min}$ and the integral is just
\begin{equation}
\begin{aligned}
	p(\mu_{\rm rel}) & = \int_{\mu_{\rm min}}^{+\infty} \diff \mu^{(1)} \; \mu^{(1)} (2 \mu_{\rm min}^{2})^2 \left(\mu^{(1)}\right)^{-3} \left(\mu_{\rm rel}\mu^{(1)}\right)^{-3} \\
	& = 4 \mu_{\rm min}^{4} \left[ \int_{\mu_{\rm min}}^{+\infty} \diff \mu^{(1)} \; \left(\mu^{(1)}\right)^{-5} \right] \mu_{\rm rel}^{-3} \\
	& = \mu_{\rm rel}^{-3}.
\end{aligned}
\end{equation}
For the latter case, $\mu^{(1)} \geq \mu_{\rm min}/\mu_{\rm rel}$ for the integral to not vanish. Therefore, the lower limit of the integral becomes $\mu_{\rm min}/\mu_{\rm rel}$ instead, and
\begin{equation}
\begin{aligned}
	p(\mu_{\rm rel}) & = \int_{\mu_{\rm min}/\mu_{\rm rel}}^{+\infty} \diff \mu^{(1)} \; \mu^{(1)} (2 \mu_{\rm min}^{2})^2 \left(\mu^{(1)}\right)^{-3} \left(\mu_{\rm rel}\mu^{(1)}\right)^{-3} \\
	& = 4 \mu_{\rm min}^{4} \left[ \int_{\mu_{\rm min}/\mu_{\rm rel}}^{+\infty} \diff \mu^{(1)} \; \left(\mu^{(1)}\right)^{-5} \right] \mu_{\rm rel}^{-3} \\
	& = 4 \mu_{\rm min}^{4} \frac{\mu_{\rm rel}^{4}}{4 \mu_{\rm min}^{4}} \mu_{\rm rel}^{-3} \\
	& = \mu_{\rm rel}.
\end{aligned}
\end{equation}
Note that this poor-man's prior for $\mu_{\rm rel}$ does not depend on the value of $\mu_{\rm min}$. Again, it can be checked against simulations. Figure \ref{fig:Poor-man} shows the results of two simulations, assuming that $\mu_{\rm rel} = 2$ and $\mu_{\rm rel} = 10$ respectively. We see that the poor-man's prior for $\mu_{\rm rel}$ is indeed independent of the value of the minimum value of the absolute magnification $\mu_{\rm min}$.

\bibliography{statistical_framework.bib}

\end{document}